\documentclass[nojss]{jss}

\usepackage{orcidlink,thumbpdf,lmodern}

\usepackage{framed}
\usepackage{amsmath, amssymb}

\newcommand{\fct}[1]{\code{#1()}}

\author{Panagiotis Papastamoulis~\orcidlink{0000-0001-9468-7613}\\Athens University of Economics and Business
   \AND Fotios S. Milienos~\orcidlink{0000-0003-1423-7132}\\Panteion University of Social and Political Sciences}
\Plainauthor{Panagiotis Papastamoulis, Fotios S. Milienos}

\title{\pkg{bayesCureRateModel}: Bayesian Cure Rate Modeling for Time to Event Data in \proglang{R}}
\Plaintitle{Bayesian Cure Rate Modeling for Time to Event Data in R}
\Shorttitle{Bayesian Cure Rate Modeling in \proglang{R}}

\Abstract{
The family of cure models provides a unique opportunity to simultaneously model both the proportion of cured subjects 
(those not facing the event of interest) and the distribution function of time-to-event for susceptibles (those facing the event). 
In practice, the application of cure models is mainly facilitated by the availability of various \proglang{R} packages. 
However, most of these packages primarily focus on the mixture or promotion time cure rate model. 
This article presents a fully Bayesian approach implemented in \proglang{R} to estimate a general family of cure rate models in the presence of covariates. 
It builds upon the work by \cite{papastamoulis2023bayesian} by additionally considering various options for describing the promotion time, including the Weibull, 
exponential, Gompertz, log-logistic and finite mixtures of gamma distributions, among others. Moreover, the user can choose any proper distribution function 
for modeling the promotion time (provided that some specific conditions are met). 
Posterior inference is carried out by constructing a Metropolis-coupled Markov chain Monte Carlo (MCMC) sampler, 
which combines Gibbs sampling for the latent cure indicators and Metropolis-Hastings steps with Langevin diffusion 
dynamics for parameter updates. The main MCMC algorithm is embedded within a parallel tempering scheme by considering 
heated versions of the target posterior distribution. The package is illustrated on a real dataset analyzing the duration 
of the first marriage considering various covariates such as the race, age and the presence of kids.}

\Plainkeywords{Metropolis-coupled MCMC, multimodal posterior distribution, time-to-event data, survival analysis, R}

\Address{
  Panagiotis Papastamoulis\\
  Department of Statistics\\
  Athens University of Economics and Business\\
  76, Patission Str.\\
  GR-10434, Athens, Greece\\
  E-mail: \email{papastamoulis@aueb.gr}\\
  URL: \url{http://www2.aueb.gr/users/papastamoulis}

  Fotios S. Milienos\\
  Department of Sociology\\
  Panteion University of Social and Political Sciences\\
  136, Syngrou Av.\\
  GR-17671, Athens, Greece\\
  E-mail: \email{milienos@panteion.gr}\\
  URL: \url{https://sites.google.com/view/fotiossmilienos}
  
}

\begin{document}




\section[Introduction: Cure Rate Models]{Introduction} \label{sec:intro}
One of the critical issues in many real-life problems is to determine whether a member of the population under study will experience a specific event or not. 
For instance, it is crucial to know the proportion of recidivism, the divorce rate, the percentage of patients fully cured of a disease, or the proportion of 
bank customers who do not default. However, studying this proportion alone is often insufficient and without providing the full picture of the problem at hand; one 
equally important aspect is the time it takes for an individual to encounter this event if it is expected to occur. 
A family of models which simultaneously allows to estimate both the proportion of \textit{cured} (those which will not experience the event of interest) 
and the distribution function of time-to-event of \textit{susceptibles} (those which will experience the event at some point) is described by 
\begin{align}\label{Sp00}
S_P(t)=p_0+(1-p_0) S(t|\text{susceptibles}),
\end{align} 
where $S_P(t)$ is the population survival function, $p_0\in[0,1]$ is the probability of being cured, also known as \textit{cure rate}  (incidence), $S(t|\text{susceptibles})$ 
is the survival function of susceptibles (latency; it is an ordinary survival function, i.e., $\lim_{t\rightarrow \infty}S(t|\text{susceptibles})=0$). 
Model \eqref{Sp00}, the \textit{mixture cure model} (e.g., \citealp{peng2021cure,Mall96,amico2018cure}, and references therein), 
has a straightforward and appealing interpretation by dividing the population into two mutually exclusive and exhaustive  groups: the cured and susceptibles. 
Therefore, using the maximum likelihood method for estimating model parameters, the cured subjects contribute with $\lim_{t\rightarrow \infty}S_P(t)=p_0$, 
and the susceptibles with $(1-p_0)f(t|\text{susceptibles})$, where $f(t|\text{susceptibles})$ is the probability density function of $S(t|\text{susceptibles})$. 

The mixture cure model is quite flexible since one may use various ways to model $S(t|\text{susceptibles})$ (parametrically, and semi/non-parametrically) or $p_0$ 
(using, for example, a logistic, probit or complementary log-log link function). There is also a competing cause approach for modeling the population distribution; suppose that there is a number of causes, $M$ (a discrete non-negative random variable), with each cause being able to deliver the event of interest, while the cured subjects are those with $M=0$ (i.e.~$p_0 = P(M = 0)$). Then, 
$$S(t|\text{susceptibles})=S(t|M>0)=\frac{P(T>t, M>0)}{1-p_0},$$
which makes model \eqref{Sp00} to be written as
\begin{align}\label{SpM0}
S_P(t)=p_0+\sum_{m=1}^{L} S_m(t) P(M=m)=\sum_{m=0}^{L} S_m(t) P(M=m),
\end{align} 
where $S_m(t)$ is the conditional survival function of subjects with $m$ causes (i.e., $S_m(t)=P(T>t|M=m)$, 
with $S_0(t)=P(T>t|M=0)=1$, for every $t$) and $M\in\{0,1,\ldots,L\}$. The mixture model may be seen as a 
special case of the competing cause cure model \eqref{SpM0}, by assuming that either  the conditional survival function $S_m(t)$ is the 
same for each $m>0$, i.e., $S_m(t)=S(t)$ for an ordinary survival function $S(t)$,  or $L=1$. 

In literature, it is typically assumed that $S_m(t)=P(T>t|M=m)=S(t)^m$, 
for some (ordinary) survival function $S(t)$, called as \textit{promotion time distribution}; this means that $S_m(t)$ is the 
minimum of a set of $m$ independent and identically distributed random variables (e.g., \citealp{Tsodikov03}). Then, \eqref{SpM0} 
becomes $S_{P}(t)=\varphi(S(t))$, where $\varphi(z)$ is the probability generating function of $M$. If $M$ follows a Poisson distribution with 
parameter $\vartheta>0$, i.e., $S_{P}(t)=\exp\{-\vartheta (1-S(t))\}=p_0^{1-S(t)}$, where	 $p_0=\exp\{-\vartheta\}$,  
we get the well known \textit{bounded cumulative hazard} or \textit{promotion time} cure model; another popular distribution of modeling the number of $M$, is the negative binomial distribution (e.g., \citealp{tournoud2007application,Castro09,pal2021simplified,koutras2017flexible}).

\cite{papastamoulis2023bayesian} provided a fully Bayesian approach for estimating the model parameters of  
\begin{align}\label{mil22}
S_P(t)=(1+\gamma \vartheta c^{\gamma \vartheta}  F(t)^\lambda)^{-1/\gamma},\gamma \in \Re,
\end{align}
with $\vartheta>0$, $\lambda>0$, $F(t)=1-S(t)$ and $c=e^{e^{-1}}$. Among the special cases of the above model are the most studied cure models, such as the promotion time ($\gamma \rightarrow 0$, $\lambda=1$), the negative binomial ($\gamma>0$, $\lambda=1$) and the mixture cure model ($\gamma=-1$, $\lambda=1$; the binomial cure model for $\gamma<0$, $\lambda=1$). Besides, the case where the population do not include cured subjects,  can also be covered, and it is not found at the boundary of the parameter space, as it is usually happens to other classes of cure models; specifically, this case is described by the scenario $(\gamma,\lambda,\vartheta)=(-1, 1, e)$ (see also \citealp{milienos2022reparameterization}). 

In the current work, we introduce the \proglang{R} \citep{R} package  \pkg{bayesCureRateModel} available from the  Compehensive R Archive Network 
at \url{https://CRAN.R-project.org/package=bayesCureRateModel}. 
The contributed package carries out a fully Bayesian approach for estimating model \eqref{mil22} under the presence 
of covariates and a (non-informative) random right censoring. To the best of our knowledge, this is the only \proglang{R} package expressly tailored for 
estimating a general family of cure rate models under a Bayesian framework.  Furthermore, it generalizes the model proposed by \cite{papastamoulis2023bayesian}, 
which used the Weibull distribution for modeling the promotion time. We extend this to model promotion time using various distributions, including Weibull, exponential, 
Gompertz, log-logistic and finite mixtures of gamma distributions. User-defined promotion time distributions are also allowed, provided that some specific conditions are met. Posterior inference is carried out by constructing 
a Metropolis-coupled Markov chain Monte Carlo sampler (MC$^3$; \citealp{altekar2004parallel}), by running in parallel various MCMC chains which target tempered versions of the posterior distribution, while allowing them to switch states. 

The rest of the paper is organized as follows. Section \ref{rpackagessec} reviews the available software for cure rate modeling. 
Section \ref{modelsoft} presents the underlying model, the Bayesian framework (Section \ref{sec:priors}), the MC$^3$ sampler (Section \ref{sec:inference}) 
and the main function of the contributed \proglang{R} package (Section \ref{sec:software}) along with the relevant methods for printing, summarizing and 
plotting the output. Section \ref{sec:illustrations} illustrates in practice our package for analysing a dataset on the duration of first marriage; this 
dataset was created by the authors, using the National Longitudinal Survey of Youth 1997 (NLSY97; \citealp{NLSY97}), a longitudinal study initiated in 1997 
that tracked a sample of American youth born between 1980 and 1984, until 2022. The article is concluded in Section \ref{sec:summary}, followed by a 
special ``Computational details'' section which discusses  parallelization and other programming issues. Technical details regarding the parameterization 
of the  distributions used throughout the paper are summarized in Appendix \ref{app:technical}. More specialized options which allow the user to define the distributional family describing the promotion time are discussed in Appendix \ref{app:user}. Two types of comparisons against alternative approaches are presented in Appendix \ref{sec:comparison}. In Appendix \ref{sec:stan} the proposed MCMC sampler is benchmarked against Hamiltonian Monte Carlo approach provided in \proglang{STAN} \citep{carpenter2017stan}. In Appendix \ref{sec:mixture}, the classification performance of the proposed model is compared against the one arising from the mixture cure rate model under the EM algorithm implementation in the \pkg{mixcure} package \citep{mixcure,peng2021cure}. 

\subsection{Software for cure rate models}\label{rpackagessec}

The application of cure models in real life problems is mainly facilitated by the availability of various \proglang{R} packages. However, most of these packages focus primarily on the mixture or promotion time cure rate model (e.g., \citealp{peng2021cure}). Next we review the available packages in \proglang{R}, devoted to cure rate modeling.

The \pkg{cuRe} package \citep{cuRe,jensen2022fitting} supports the parameter estimation for both the mixture and the promotion time cure model using either a parametric approach (e.g., assuming a logit function for the cure rate and a Weibull distribution for the survival function of susceptibles) or a spline-based formulation. The \pkg{smcure} package \citep{smcure, smcurepaper} employs the Expectation-Maximization (EM) algorithm \citep{dempster1977maximum} for estimating the mixture cure model, assuming either the proportional hazard or the accelerated failure time model for the latency. The proportional hazard mixture cure model is also treated by the \pkg{geecure} package \citep{niu2018geecure,niu2014marginal}, which uses generalized estimating equations and a modified EM algorithm to estimate model parameters; it is also suitable for analyzing clustered survival data. The \pkg{mixcure} package \citep{mixcure,peng2021cure} gathers a set of parametric and semiparametric approaches from existing \proglang{R} packages for studying the mixture cure model. The \pkg{flexsurvcure} package \citep{flexsurvcure} fits the mixture or the promotion time cure model parametrically, while the \pkg{spduration} \citep{spduration} and \pkg{EventPredInCure} packages \citep{EventPredInCure, chen2016predicting} also provide options for fitting parametrically the mixture cure models. The \pkg{GORcure} package \citep{GORcure} refers to a flexible mixture cure model, with the proportional hazard mixture cure model and the proportional odds mixture cure model as special cases, and handles interval-censored data, as well.

A non-parametric approach for estimating both the incidence and latency, under the mixture cure model, is provided by the \pkg{npcure} package \citep{npcure,lopez2024npcure} (using one continuous covariate); a non-parametric method for the mixture cure model is also adopted in the \pkg{npcurePK} package \citep{npcurePK}, when the cure status is partially known. The \pkg{curephEM} package \citep{curephEM, hou2018nonparametric} offers a non-parametric maximum likelihood approach for the mixture cure rate model, where the cure status of some subjects may also be assumed known.

Other significant aspects of cure modeling are further studied in various packages, such as, a) the  \pkg{CureAuxSP} \citep{CureAuxSP,ding2024cureauxsp}, wherein the use of auxiliary subgroup survival probabilities provided by external sources are incorporated into the mixture cure model estimation, b) the \pkg{CureDepCens} \citep{CureDepCens,schneider2022free} which considers the case of dependent censoring under the promotion time cure model, c) the \pkg{hdcuremodels} \citep{hdcuremodels, fu2022controlled} which accounts for high-dimensional data under the mixture cure model, 
d)  the \pkg{penPHcure} \citep{penPHcure} focusing on the variable selection problem, under the proportional hazard mixture cure model with time-varying covariates, e) 
the \pkg{thregI }\citep{thregI} treating a threshold regression concept under the mixture cure rate model, f) the \pkg{miCoPTCM} \citep{miCoPTCM} where the promotion time cure model, with mis-measured covariates is considered, and e) the \pkg{NPHMC} \citep{NPHMC} for computing sample size under the proportional hazard mixture cure model. 

The \pkg{rstpm2} package \citep{rstpm2,jakobsen2020generalized,jensen2022fitting} refers to a class of models known as latent cure models, which under specific assumptions, permit the estimation of cure rate and survival function of susceptibles using a common set of parameters. The \pkg{nltm} package \citep{nltm, tsodikov2003semiparametric,Tsodikov03} deals with, among other things, the class of non-linear transformation cure models. The class of power series cure models are studied by the \pkg{PScr} \citep{PScr}, adopting the EM algorithm for parameter estimation.

Apart from the \proglang{R} packages mentioned above, there are also few options for cure rate modeling, available in other statistical software such as \proglang{SAS} \citep{sas2024} and \proglang{Stata} \citep{stata2023}. To be more specific, for the mixture cure model, one could follow a parametric and semi-parametric approach provided by the  \proglang{SAS}  macro \pkg{PSPMCM} \citep{corbiere2007sas}, or a fairtly specification approach, based on \proglang{SAS} macro \pkg{proc NLMIXED} and the code provided by \cite{rondeau2013cure}. \proglang{Stata} module \pkg{CUREREGR} also fits the mixture or promotion time cure model adopted a parametric approach \citep{buxton2013cureregr}; see also \cite{lambert2007modeling} for the modules \pkg{STRSMIX} and
\pkg{STRSNMIX}, or \cite{crowther2014stgenreg}, for  \pkg{STGENREG} (see also \cite{crowther2020merlin}). There is also the Microsoft Windows application \pkg{CANSURV} \citep{yu2005cansurv}, for fitting a parametric mixture cure model. 
 
Evidently, these programs collectively enhance the application of cure models in data analysis by offering a range of methods and approaches to estimate cure rates and model survival functions. 
However, it appears that Bayesian methods for cure models are scarcely represented among the available \proglang{R} packages, if at all. The \pkg{BayesSPsurv} package \citep{bolte2021bayesspsurv}  fits Bayesian cure rate survival models with time-varying covariates, taking into account spatial autocorrelations, considering the Weibull and log-logistic distributions (we note however that it has been removed from CRAN on 2023-06-14). Of course, one could try ``MCMC on the autopilot'' software such as \proglang{STAN} \citep{carpenter2017stan}, \proglang{NIMBLE} \citep{doi:10.1080/10618600.2016.1172487} or \proglang{BUGS}/\proglang{WinBugs} \citep{lunn2000winbugs, ntzoufras2011bayesian}. 
However, the potential multimodality \citep{papastamoulis2023bayesian} of the posterior distribution of cure rate models  would make these implementations prone to 
converging to minor modes and produce sub-optimal inferences (see Section \ref{sec:stan} in the Appendix). In addition, discrete parameters are not allowed in \proglang{STAN}, therefore inference for the 
latent cured status would not be straightforward. Finally, \proglang{STAN}, \proglang{NIMBLE} and \proglang{WinBUGS} require a certain level of statistical and programming knowledge and users 
must still understand model specification, convergence diagnostics, and interpretation of results.

\section{Model, inference and software}\label{modelsoft}

Denoting by $C_i$ and $T_i$ the censoring time and time-to-event of the $i$-th subject, respectively, our observed data consist of $Y_i = \min\{T_i,C_i\}$, along 
with the censoring  indicator. It is necessary to mention, that under our scenario the cured subjects never failed and then, censoring times corresponds to them.  
Analytically, let $\boldsymbol{y} = (y_1,\ldots,y_n)$ denote the observed data, which correspond to time-to-event or censoring time, and $\boldsymbol{x}_i = (x_{i1},\ldots,x_{ik})'$ being the vector of $k$ covariates, for subject $i=1,\ldots,n$ (wherein $x_{i1}$ may correspond to a constant term, and thus $x_{i1}=1$, or not). These covariates affect the population  survival function \eqref{mil22} through $\vartheta$ assuming an exponential link function $\vartheta(\boldsymbol x)=\exp\{\boldsymbol{x}\boldsymbol{\beta}'\}$, thus \eqref{mil22} is expressed as 
\begin{align}\label{eq0001}
S_P(y_i;\boldsymbol\theta) = \left(1 + \gamma\exp\{\boldsymbol{x}_i\boldsymbol{\beta}'\}c^{\gamma\exp\{\boldsymbol{x}_i\boldsymbol{\beta}'\}}F(y;\boldsymbol\alpha)^\lambda\right)^{-1/\gamma},\quad i = 1,\ldots,n,
\end{align}
where $F(\cdot,\boldsymbol\alpha)$  denotes the promotion time cumulative distribution function parameterized by a generic parameter $\boldsymbol\alpha\in\mathcal A$. It is necessary to mention that due to identifiability issues, we must assume the existence of a continuous covariate with
non-negligible effect on $\vartheta$ (see also \citealp{papastamoulis2023bayesian, milienos2022reparameterization}). Hence, the parameter vector $\boldsymbol\theta$ is decomposed as $\boldsymbol\theta = (\boldsymbol\alpha', \boldsymbol\beta', \gamma,\lambda)$, where  $\boldsymbol\alpha$ are the parameters of the promotion time distribution, 
$\boldsymbol\beta\in\mathbb{R}^{k}$ are the regression coefficients, 
$\gamma\in\mathbb R$ and $\lambda > 0$. The cure rate inferred from model \eqref{eq0001} is given by 
\[\lim_{y\rightarrow \infty}S_P(y;\boldsymbol\theta)=p_0(\boldsymbol{x}_i;\boldsymbol{\theta}) = \left(1 + \gamma\exp\{\boldsymbol{x}_i\boldsymbol{\beta}'\}c^{\gamma\exp\{\boldsymbol{x}_i\boldsymbol{\beta}'\}}\right)^{-1/\gamma}.\]

Assuming that the $n$ observations are independent, the observed likelihood function is defined as
\[
L=L({\boldsymbol \theta}; {\boldsymbol y}, {\boldsymbol x})=\prod_{i=1}^{n}f_P(y_i;{\boldsymbol\theta},{\boldsymbol x}_i)^{\delta_i}S_P(y_i;{\boldsymbol \theta},{\boldsymbol x}_i)^{1-\delta_i},
\]
where $f_P(y;\boldsymbol\theta)$ is the population probability density function, namely 
$f_P(y;\boldsymbol\theta)=-\frac{\partial S_P(y;\boldsymbol\theta)}{\partial y}$,  
while $\delta_i=1$ if the $i$-th observation corresponds to time-to-event, and $\delta_i=0$ otherwise (i.e., for a censoring time).

The promotion time distribution can be a member of standard families, specifically,   the Weibull, gamma, Lomax, Gompertz, and log-logistic distribution,  and in such a case $\alpha = (\alpha_1,\alpha_2)\in (0,\infty)^2$. Also considered is the exponential distribution with one parameter $\alpha_1\in (0,\infty)$ and  Dagum distribution, which has three parameters $(\alpha_1,\alpha_2,\alpha_3)\in(0,\infty)^3$ (see Appendix \ref{app:technical} for the parameterization of these distributions). 

If the previous parametric assumptions are not justified, the promotion time can be user-defined, as long as it is a valid univariate distribution function $f(y;\boldsymbol{\ddot\alpha})$, with $y > 0$, parameterized by $\boldsymbol{\ddot{\alpha}} = (\ddot{\alpha}_{1}, \ldots,\ddot{\alpha}_{d})\in (0,\infty)^d$, that is, only positive parameters are allowed. Moreover, one can assume that the promotion time distribution is a finite mixture of distributions of the form 
\begin{equation}
\label{eq:mixture}
\sum_{i=1}^{K}\dot{\alpha}_i f(\cdot;\boldsymbol{\ddot\alpha}_i),
\end{equation}
 where  $\dot\alpha_i>0$, $\boldsymbol{\ddot{\alpha}}_i\in (0,\infty)^d$ for  $i=1,\ldots,K$ and $\sum_{i=1}^{K}\dot{\alpha}_i = 1$. Hence, the parameter vector $\boldsymbol \alpha$ of the promotion time distribution is now written as $\boldsymbol \alpha = (\boldsymbol{\dot{\alpha}}, \boldsymbol{\ddot{\alpha}})$. Our package has a built-in function for fitting finite mixtures of Gamma distributions, however, the user can define arbirtrary univariate finite mixture models as long as the distribution describing each component has strictly positive parameters as in Equation \eqref{eq:mixture}. The number of the mixture components can be selected according to information criteria such as the Bayesian Information Criterion (BIC; \citealp{schwarz1978estimating}). The reader is referred to Appendix \ref{app:user} for a practical illustration using finite mixtures of log-normal distributions. 

The binary vector $\boldsymbol{I} = (I_1,\ldots,I_n)$ contains the (latent) cure indicators, that is, $I_i = 1$ if the $i$-th subject is susceptible and $I_i = 0$ if the $i$-th subject is cured.  $\Delta_0$ denotes the subset of $\{1,\ldots,n\}$ containing the censored subjects, whereas $\Delta_1 = \Delta_0^c$ is the (complementary) subset of uncensored subjects. 
The complete likelihood of the model is 
\begin{align}
\label{eq:cll}
L_c(\boldsymbol{\theta};\boldsymbol{y}, \boldsymbol{I}) &= \prod_{i\in\Delta_1}(1-p_0(\boldsymbol{x}_i,\boldsymbol\theta))f_U(y_i;\boldsymbol\theta,\boldsymbol{x}_i)\prod_{i\in\Delta_0}p_0(\boldsymbol{x}_i,\boldsymbol\theta)^{1-I_i}\{(1-p_0(\boldsymbol{x}_i,\boldsymbol\theta))S_U(y_i;\boldsymbol\theta,\boldsymbol{x}_i)\}^{I_i},
\end{align}
with $f_U(y_i;\boldsymbol\theta,{\boldsymbol x}_i)=\frac{f_P(y_i;\boldsymbol\theta,{\boldsymbol x}_i)}{1-p_0({\boldsymbol x}_i;\boldsymbol\theta)}$ and $S_U(y_i;\boldsymbol\theta,{\boldsymbol x}_i)=\frac{S_P(y_i;\boldsymbol{\theta},{\boldsymbol x}_i)-p_0({\boldsymbol x}_i;\boldsymbol\theta)}{1-p_0({\boldsymbol x}_i;\boldsymbol\theta)}$ denoting the probability density and survival function of the susceptibles, respectively. It is worth noting that the set of covariates influences both the cure rate and the distribution function of the susceptible individuals. This is inherent to the adopted approach of incorporating covariate effects through the parameter $\mathcal{\theta}$.


\subsection{Prior assumptions}\label{sec:priors}

The prior assumptions are similar to the ones introduced in \cite{papastamoulis2023bayesian}. In brief, we assume that the prior distribution factorizes as follows
\begin{equation}
\label{eq:prior}
\pi(\boldsymbol\theta) = \pi(\boldsymbol\alpha)\pi(\boldsymbol\beta)\pi(\gamma)\pi(\lambda),
\end{equation}
where the joint prior distribution of the regression coefficients is
\begin{equation}
\label{eq:beta_prior}
\pi(\boldsymbol\beta) = \mathcal N(\boldsymbol\mu,\boldsymbol\Sigma),
\end{equation}
with $\mathcal N(\boldsymbol\mu,\boldsymbol\Sigma)$ denoting the multivariate Normal distribution, with fixed mean $\boldsymbol\mu\in\mathbb R^{k}$ and $\Sigma$ a fixed $k\times k$ positive definite matrix, where $k$ denotes the number of columns in the design matrix. The default values are set to $\boldsymbol\mu = (0,\ldots,0)^\top$, while $\Sigma = 100 \mathcal I_{k}$, with $\mathcal I_k$ denoting the $k\times k$ identity matrix. 

For parameter $\gamma$,  we assume that 
\begin{equation}
\label{gamma_prior}
\pi(\gamma) = 
\frac{b_\gamma^{a_\gamma}}{2\Gamma(a_\gamma)}|\gamma|^{a_\gamma-1}\exp\{-b_\gamma|\gamma|\}\mathrm{I}_{\mathbb R}(\gamma),
\end{equation}
where $a_\gamma>0$ and $b_\gamma>0$ denote fixed hyper-parameters. The default parameters are set to $a_\gamma=b_\gamma=1$ which reduces \eqref{gamma_prior} to a standard Laplace distribution. 

The parameter $\lambda$ follows an inverse gamma distribution,
\begin{equation}
\label{lambda_prior}
\pi(\lambda) = \mathcal{IG}(a_\lambda,b_\lambda),
\end{equation}
where $a_\lambda > 0$ and $b_\lambda > 0$ fixed hyper-parameters. The default values are set to $a_\lambda = 2.1$ and $b_\lambda  = 1.1$. 
 
In all cases excluding finite mixtures of distributions,   $\pi(\boldsymbol\alpha)$ consists of a product of independent inverse gamma distributions, one for each element of the vector $\alpha = (\alpha_1,\ldots,\alpha_d)'$, i.e., 
\begin{equation}
\label{eq:alpha_prior}
\pi(\boldsymbol\alpha) = \prod_{j=1}^{d}\mathcal{IG}(a_{j}, b_{j}),
\end{equation}
where $a_{j}$ and $b_{j}$ are positive fixed hyper-parameters. 

In the case of the finite mixture model in Equation \eqref{eq:mixture}, the prior distribution for $\boldsymbol \alpha = (\boldsymbol{\dot{\alpha}}, \boldsymbol{\ddot{\alpha}})$ factorizes into a product of independent inverse gamma distributions for the component-specific parameters of the mixture components, as well as a term which corresponds to a Dirichlet prior on the mixing proportions $\boldsymbol{\dot{\alpha}}$. Hence,
\begin{equation}
\label{eq:alpha_prior2}
\pi(\boldsymbol{\dot\alpha}, \boldsymbol{\ddot{\boldsymbol{\alpha}}}) = \pi(\boldsymbol{\dot{\alpha}})\pi(\boldsymbol{\ddot{\alpha}}) = \prod_{i=1}^{K}\prod_{j=1}^{d}\mathcal{IG}(a_{ij}, b_{ij})\mathcal{D}(\alpha_0,\ldots,\alpha_0),
\end{equation}
where $\mathcal{D}(\alpha_0,\ldots,\alpha_0)$ denotes the Dirichlet distribution with common concentration parameter $\alpha_0>0$ (fixed hyper-parameter). The default values are: $a_{ij} = 2.1$, $b_{ij} = 1.1$ for $i=1,\ldots,K$; $j=1,\ldots,d$ and $\alpha_0 = 1$. 

We note that these default choices of the hyper-parameters are aligned to the ``regularized'' prior setup in the paper of \cite{papastamoulis2023bayesian}. We also refer the reader to \cite{papastamoulis2023bayesian} for various sensitivity checks regarding the prior set-up against vague prior distributons.  

\subsection{MCMC inference}\label{sec:inference}

Under our Bayesian setup, inference is based on the joint posterior distribution of model parameters and latent cured status indicators
\begin{align*}
\pi(\boldsymbol\theta,\boldsymbol I|\boldsymbol y, \boldsymbol x)&\propto f(\boldsymbol y|\boldsymbol\theta,\boldsymbol I, \boldsymbol x)\pi(\boldsymbol I,\boldsymbol\theta)= f(\boldsymbol y, \boldsymbol I|\boldsymbol\theta, \boldsymbol x)\pi(\boldsymbol\theta)\\
&\propto L_c(\boldsymbol\theta;\boldsymbol y,\boldsymbol I)\pi(\boldsymbol\theta),
\end{align*}
where in the first line of the previous Equation we have used the generic notation $f(x|y)$ to refer to the conditional distribution of $x$ given $y$, while 
$L_c$ denotes the complete log-likelihood defined in Equation \eqref{eq:cll} and $\pi(\boldsymbol\theta)$ is the prior distribution in Equation \eqref{eq:prior}. Naturally, $\pi(\boldsymbol\theta,\boldsymbol I|\boldsymbol y, \boldsymbol x)$ is  intractable hence we use Markov chain Monte Carlo in order to approximate quantities of interest via  simulations. 

The latent cured status indicators ($\boldsymbol I$) are updated by performing a Gibbs step, i.e.~simulating from the full conditional posterior distribution $\pi(\boldsymbol I|\boldsymbol\theta,\boldsymbol y,\boldsymbol x)$.  The components of the parameter vector $\boldsymbol\theta$ are updated using a combination of Metropolis-within-Gibbs step or a Metropolis-Adjusted Langevin diffusion \citep{10.2307/3318418, https://doi.org/10.1111/1467-9868.00123, girolami2011riemann} (MALA) step. The difference in these two alternatives is that the first one proposes small changes at each component of the parameter vector sequentially, while the second proposes to update all components simultaneously taking into account information from the gradient of the logarithm of the conditional posterior distribution $\pi(\boldsymbol\theta|\boldsymbol y, \boldsymbol x, \boldsymbol I)$. 

Let $\theta_j$ represents a univariate component of the parameter vector $\boldsymbol\theta$, so that $\boldsymbol\theta = (\theta_1,\ldots,\theta_{k+d+2})$. We also denote by $\tilde\theta_j$ the candidate state of $\theta_j$, $j=1,\ldots,k+d+2$. Single-site updates are attempted in a Metropolis-within-Gibbs step, using log-normal proposal distributions of the form 
\begin{equation}
\label{eq:single_site_update}
\log\tilde\theta_j = \log\theta_j + \varepsilon_{j},
\end{equation}
where $\varepsilon_{j}\sim\mathcal{N}(0,\sigma^2_j)$ and $\sigma_j>0$ is a fixed scale parameter of the proposal distribution. This is directly applicable for all parameters in our model, excluding the case where the promotion time is a mixture of gamma distributions due to restricted parameter space of mixing proportions. In this case we follow the reparameterization strategy suggested in \cite{MARIN2005459} and apply the previous proposal mechanism. 

The MALA step generates a candidate state $\boldsymbol{\tilde\theta}$ as follows
\begin{equation}
\label{eq:proposal}
\tilde{\boldsymbol\theta} = \boldsymbol\theta + \tau\nabla\log \pi(\boldsymbol\theta|\boldsymbol y, \boldsymbol x, \boldsymbol I) + \sqrt{2\tau}\boldsymbol\varepsilon, 
\end{equation}
where $\tau>0$ (fixed parameter which determines the scale of the proposal distribution), $\boldsymbol\varepsilon\sim\mathcal N(\boldsymbol 0, \mathcal I_{k+d+2})$ and $\nabla\log \pi(\boldsymbol\theta|\boldsymbol y, \boldsymbol x, \boldsymbol I)$ denotes the gradient vector of the logarithm of the full conditional posterior distribution of $\boldsymbol\theta$. 

\cite{papastamoulis2023bayesian} demonstrated that the posterior distribution of this model can exhibit multiple minor modes. 
In such cases typical MCMC samplers can become trapped into one minor mode and/or exhibit poor convergence properties. In order to efficiently sample from the joint posterior distribution of the model, the basic MCMC sampler is embedded within a parallel tempering scheme which allows tempered chains to interact according to the so-called Metropolis-coupled MCMC ($\mbox{MC}^{3}$) \citep{geyer1991markov, geyer1995annealing, altekar2004parallel} strategy. We consider a series of $C\geqslant 2$ MCMC chains, each one targetting a ``heated'' version of the original posterior distribution, that is,
\begin{equation}
\label{eq:heated_post}
\pi_c(\boldsymbol\theta,\boldsymbol I)\propto\pi(\boldsymbol\theta,\boldsymbol I|\boldsymbol y,\boldsymbol x)^{h_c} \propto f(\boldsymbol y, \boldsymbol I| \boldsymbol\theta, \boldsymbol x)^{h_c}\pi(\boldsymbol\theta)^{h_c},\quad c=1,\ldots,C,
\end{equation}
where $0\leqslant h_c \leqslant 1$ is a given constant corresponding to the temperature of chain $c$. Note that when  raising a distribution to a power between $0$ and $1$ it becomes flatter, thus, easier to explore. The sequence of temperatures is such that $h_1\geqslant\ldots\geqslant h_C$ with $h_1 = 1$ (corresponding to the target posterior distribution).  
The $C$ chains which target the heated posterior distributions are allowed to interact by proposing swaps between pairs of (adjacent) chains after a small number  of usual MCMC iterations, referred to as \textit{sweeps}. The completion of the pre-defined number of sweeps (typical values are 5 or 10 sweeps) is referred to as an \textit{MCMC cycle}. At the end of each MCMC cycle, a swap move attempts to switch the values of two randomly proposed (adjacent) chains. The proposed swaps are accepted with the usual Metropolis-Hastings acceptance probability. Effectively, accepted swaps enhance the ability of the sampler to freely explore the posterior surface. For more details the reader is referred to Algorithm 1 and 2 in  \cite{papastamoulis2023bayesian}.

Finally, we are also producing a list of ``discoveries'', that is, subjects in the sample that are deemed as ``cured'' after controlling the False Discovery Rate \citep{benjamini1995controlling} at a desired level $0<\alpha<1$. This is doable since our MCMC sampler produces an estimate of the posterior  ``cured'' probability, i.e.~$\mathrm{P}(I_i = 0|\boldsymbol y, \boldsymbol x)$, for each subject $i=1,\ldots,n$. The reader is referred to Section 4 in \cite{papastamoulis2023bayesian} for details (see also \citealp{papastamoulis2018bayesian}). 

\subsection{Software}\label{sec:software}
The main function of the \pkg{bayesCureRateModel} package is \fct{cure\_rate\_MC3} and the accompanying \fct{print}, \fct{summary}, \fct{predict} and \fct{plot} methods. Its most 
important arguments are
\begin{Code}
cure_rate_MC3(formula, data, nChains, mcmc_cycles, alpha, nCores, 
  promotion_time, ...),
\end{Code}
where \code{formula} is an object of class \code{"formula"}, i.e.~a symbolic description of the model to be fitted. Then, the left hand side of the formula 
should correspond to a \code{Surv} object, a class inherited from the \pkg{survival} package \citep{survival, survival-book}. 
Assume, for instance, that \code{time} is the variable containing the observed (possibly censored) times and \code{censoring} is a binary vector 
corresponding to censoring indicators (1 for time-to-event entries, and 0 for censored). Then, the left hand side of the formula 
should be defined as \code{Surv(time, censoring)}, while any covariates (which affect parameter $\vartheta$, through the exponential link function) are given in the right 
hand side (e.g., \code{Surv(time, censoring)~x1+x2}). The argument \code{data} should be a data frame containing all variable names included in \code{formula}.

The number of MCMC cycles must be provided to \code{mcmc_cycles}. The \code{nChains} is a positive integer corresponding to the number of heated chains in the $\mathrm{MC}^3$ scheme, that is, $C$ in Equation \eqref{eq:heated_post}.
The \code{alpha} argument is a decreasing sequence $(h_1,\ldots,h_C)$ in $[1,0]$ of \code{nChains} temperatures, see Equation \eqref{eq:heated_post}. The first value should always be equal to 1, which corresponds to the target posterior distribution (that is, the first chain). The default values are set as
\[
h_c = \frac{1}{(1+\varepsilon_0)^{c^{d_0} - 1}},\quad c=1,\ldots,C,
\]
where $\varepsilon_0 > 0$, $d_0 > 0$ and $C = 12$. We have used $\varepsilon_0=0.001$ and 
$$
d_0 = \begin{cases}5&, \mbox{if}\quad C \leqslant 4\\
3.5&, \mbox{if}\quad 5 \leqslant C \leqslant 8\\
3&, \mbox{if}\quad  C \geqslant 9
\end{cases}.
$$

The \code{nCores} argument corresponds to the number of cores used for parallel processing. In case where \code{nCores = 1} the computation is done on a single core. When setting \code{nCores} $>1$, the \code{nChains} heated chains are processed in parallel using \code{nCores} workers. Obviously, it should hold that \code{nCores} $\leqslant$ \code{nChains}. Parallelization is recommended in Unix-like systems (e.g.~Linux, MacOS), however it is not suggested in Windows: see the discussion in the special ``Computational details'' section. 

The \code{promotion_time} argument defines details of the parametric family of distribution describing the promotion time and corresponding prior distributions. It should be a list containing the following entries 
\begin{itemize}
\item[]     \code{family}: Character string specifying the family of
distributions describing the promotion time. The available options are: \code{"exponential"}, \code{"weibull"}, \code{"gamma"}, \code{"logLogistic"}, \code{"gompertz"}, 
\code{"gamma_mixture"}, \code{"lomax"} and \code{"dagum"}. If not provided, it will be set to \code{weibull} by default. Also available are the options 
\code{"user"} and \code{"user_mixture"} which allow the user to define their own promotion time distribution family, or a 
finite mixture of a given family of distributions, respectively (see Appendix \ref{app:user} for some examples).
\item[] \code{prior_parameters}:  Values of hyper-parameters of the Inverse Gamma prior
          distributions of the parameters $\boldsymbol{\alpha}$, see Equation \eqref{eq:alpha_prior}. If not provided by the user, the default values are being used.  It should correspond to a $d\times 2$-dimensional matrix, where $d$ denotes the number of parameters in \code{family}, in all cases besides \code{family = "gamma_mixture"} or \code{family = "user_mixture"}. In the latter cases, \code{prior_parameters} corresponds to a $d\times 2\times K$-dimensional array, where $K$ denotes the number of mixture components. All entries should be non-negative.
\item[] \code{prop_scale}: The scale of the proposal distributions (see \eqref{eq:single_site_update}) for each
          parameter in $\boldsymbol \alpha$. If not provided, the default values 
		  are set equal to a $d$-dimensional vector with all values equal to $0.1$.
\item[] \code{dirichlet_concentration_parameter}: Relevant only in the case of
          the \code{family = "gamma_mixture"} or \code{family = "user_mixture"}. Positive scalar 
          determining the (common) concentration parameter of the
          Dirichlet prior distribution of mixing proportions in Equation \eqref{eq:alpha_prior2}. If not provided by the user, the default value of 1 is being used.
\end{itemize}

Further arguments regarding the hyper-parameters of the prior distribution to \fct{cure\_rate\_MC3} are the following. The arguments \code{a_g} and \code{b_g} correspond to the hyper-parameters $a_{\gamma}$ and $b_\gamma$, respectively, of the prior distribution for $\gamma$ in Equation \eqref{gamma_prior}. The arguments \code{mu_b} and \code{Sigma} correspond to the hyper-parameters $\boldsymbol\mu$ and $\Sigma$, respectively, of the multivariate Normal prior distribution for $\boldsymbol\beta$ in Equation \eqref{eq:beta_prior}. The arguments \code{a_l} and \code{b_l} correspond to the hyper-parameters $a_{\lambda}$ and $b_\lambda$, respectively, of the prior distribution for $\lambda$ in Equation \eqref{lambda_prior}.

Other parameters that control the sampler are the following. The \code{g_prop_sd}, \code{b_prop_sd} and \code{lambda_prop_scale} arguments denote positive constants corresponding to the scales for the proposal distribution of the standard single-site  Metropolis-Hastings updates for $\gamma$, $\boldsymbol\mu$ and $\lambda$, respectively, in Equation \eqref{eq:single_site_update}. The \code{tau_mala} denotes the positive scalar corresponding to the scale $(\tau)$ of the MALA proposal in Equation \eqref{eq:proposal}. In each step of the MCMC sampler, the MALA proposal is attempted with probability corresponding to \code{mala} argument. Whenever setting  \code{mala} to a positive value strictly smaller than 1, the sampler will perform Metropolis-Hastings updates with probability 1 - \code{mala}. In such a case, the argument \code{single_MH_in_f} denotes the probability for attempting a series of single site  updates and with probability 1 - \code{single\_MH\_in\_f} a Metropolis-Hastings move will attempt to simultaneously update all continuous parameters.

 The \fct{cure\_rate\_MC3} function returns an object of class \code{bayesCureModel}, containing the MCMC sample among other quantities of interest. More specifically, an object of class \code{bayesCureModel}, is a list with the following entries
\begin{itemize}
\item \code{mcmc_sample}: Object of class \code{mcmc} (see the \pkg{coda} package), containing the generated MCMC sample for 
the target chain. The column names correspond to: 
\begin{itemize}
\item \code{g_mcmc}:  Sampled values for parameter $\gamma$.
\item \code{lambda_mcmc}:  Sampled values for parameter $\lambda$.
\item \code{alpha1_mcmc} $...$ \code{alphad_mcmc}:  Sampled values for parameter $\alpha_1,\ldots,\alpha_d$ of the promotion time distribution $F(\cdot;\alpha_1,\ldots,\alpha_d)$ 
in Equation \eqref{eq0001} where $d$ depends on the family used in \code{promotion_time}.
\item \code{b0_mcmc} $\ldots$ \code{bk_mcmc}:  Sampled values for the regression coefficients, depending on the design matrix of the model.
\end{itemize}
\item \code{latent_status_censored}:  A data frame with the simulated binary latent status for each censored item. 
\item \code{complete_log_likelihood}: The complete log-likelihood for the target chain.
\item \code{swap_accept_rate}: The acceptance rate of proposed swappings between adjacent MCMC chains.
\item \code{all_cll_values}:  The complete log-likelihood for all chains.
\item \code{input_data_and_model_prior}: A list containing the input data, model specification and prior parameters values.
\item \code{log_posterior}: The logarithm of the (non-augmented) posterior distribution (after integrating the latent cured-status parameters out), up to a normalizing constant.
\item \code{map_estimate}: The Maximum A Posterior estimate of parameters.
\item \code{BIC}: Bayesian Information Criterion of the fitted model.
\item \code{AIC}: Akaike Information Criterion of the fitted model.
\item \code{residuals}: The Cox-Snell residuals of the fitted model.
\item \code{initial_values}: The starting values per chain.
\end{itemize}

The \fct{print} method returns a synopsis of the fitted model including information criteria and the Maximum A Posteriori (MAP) estimate of the parameters arising from the joint posterior distribution, i.e.~the MCMC analogue of $\boldsymbol{\theta}^{\mathrm{MAP}}=\mathrm{argmax}_{\boldsymbol\theta}\pi(\boldsymbol\theta|\boldsymbol y,\boldsymbol x)$. The main reason for reporting the MAP estimate is due to the fact that the posterior distribution may exhibit minor modes \citep{papastamoulis2023bayesian}, thus, other summaries such as the posterior means (typically used in Bayesian inference) may not make sense.  In any case, the user can conveniently retrieve them since the MCMC output is returned as an \code{mcmc} object, a class inherited from the \pkg{coda} package \citep{coda}. We should also mention here that \cite{papastamoulis2023bayesian} demonstrated via extended simulation studies that the MAP estimates arising from the proposed methodology are more accurate than Maximum Likelihood estimates arising from the Expectation-Maximization algorithm.

More detailed summaries are provided by the \fct{summary} method, including Highest (posterior) Density Intervals for each parameter and a list of cured items in the sample (if any) when controlling the FDR at a desired level (see last paragraph of Section \ref{sec:inference} for details). Also, it is used to post-process the MCMC draws in order to compute the survival function and the conditional cured probability for specific covariate levels. 

The \fct{plot} method can be used to visualize the estimated marginal posterior distribution of each parameter, the survival function or the 
conditional cured probability for specific covariate levels, along with credible intervals. Finally, the \fct{residuals} and \fct{predict} methods 
return the Cox-Snell residuals (\citealp{cox1968general}) of the fitted model and predicted values (survival function, cured probability, hazard and 
cumulative hazard rate), respectively. The details of the implementation will be clarified in the next section.



\section{Illustrations} \label{sec:illustrations}
We illustrate our method using a dataset which is incorporated in our package. This dataset was created by the authors, using the National Longitudinal Survey of Youth 1997 (NLSY97) which is a longitudinal study, tracking a sample of American youth born between 1980 and 1984 (\citealp{NLSY97}). Starting in 1997, 8984 participants, aged 12 to 17 at the time, were first interviewed. This cohort has been surveyed 20 times so far, with biennial interviews now in place; hence, data consists of Round 1 (1997-98) through Round 20 (2021-2022). The event of interest in our analysis is whether a participant's first marriage ended or not; therefore, the time-to-divorce (in years) of the first marriage and whether it is actually an event time (divorce) or a censored time are our primary variables. 

Of the 8984 participants, we found that 5029 have been married at least once. However, excluding some cases due to missing information (making us unable to compute the time-to-event, or the covariate values), we came up with 3956 participants. The set of covariates consists of: age of respondent (in years) at the time of first marriage, whether there were kids during the first marriage (\code{"yes"}) or not (\code{"no"}), and race of respondent with levels corresponding to: \code{"black"}, \code{"hispanic"} and \code{"other"}. A sample of $n = 1500$ participants from the previously mentioned group was ultimately included in our package's dataset by choosing random samples of 500 individuals from each race. The following chunk loads the dataset.
\begin{Schunk}
\begin{Sinput}
R> library("bayesCureRateModel")
R> library("survival")
R> data(marriage_dataset)
R> str(marriage_dataset, strict.width = 'cut')
\end{Sinput}
\begin{Soutput}
'data.frame':	1500 obs. of  6 variables:
 $ id       : num  8885 7307 7180 5806 6879 ...
 $ censoring: num  0 0 0 0 1 0 0 0 1 0 ...
 $ time     : num  17.25 4.92 13.25 9.33 2.58 ...
 $ age      : num [1:1500, 1] -0.31865 1.5671 0.37691 0.26872 -0.009..
  ..- attr(*, "scaled:center")= num 26.6
  ..- attr(*, "scaled:scale")= num 5.39
 $ kids     : Factor w/ 2 levels "no","yes": 2 2 1 2 2 1 2 2 2 2 ...
 $ race     : Factor w/ 3 levels "black","hispanic",..: 1 1 1 1 1 1 ..
\end{Soutput}
\end{Schunk}
There are 1018 censored items and the remaining 482 observations constitute time-to-events (divorce). The data frame \code{marriage_dataset} contains the recorded (event or censoring) time ($\boldsymbol y$) data in column \code{time}, the censoring status ($\boldsymbol\delta$) in column \code{censoring}, the (standardized) continuous covariate (\code{age}) and two factor covariates in columns \code{kids} and \code{race}. Note that we have also loaded the \pkg{survival} package in order to define the response variable as \code{Surv} object in the code snippet below. The interpretation of a cure rate is supported by a long-term follow-up, approximately 20 years in average, with the Kaplan-Meier curve showing a sustained plateau, albeit not fully definitive; see e.g.~\cite{othus2020bias, selukar2023receus, xie2024testing}). At first, we fit the basic exponential model using 4 tempered chains running on a single core for a total of 15000 MCMC cycles. 
\begin{Schunk}
\begin{Sinput}
R> mcmc_cycles <- 15000; nChains <- 4; nCores <- 1
R> set.seed(10, kind = "L'Ecuyer-CMRG")
R> run_exp <- cure_rate_MC3(Surv(time, censoring) ~ age + kids + race,
+    data = marriage_dataset, nChains = nChains, mcmc_cycles = mcmc_cycles,  
+    nCores = nCores, promotion_time = list(family = 'exponential'), 
+    verbose = FALSE)
\end{Sinput}
\begin{Soutput}
20 MCMC cycles required 1.11 secs. Expect a total run-time of: 833.5 secs. 
\end{Soutput}
\end{Schunk}
Next, we consider a Weibull model. 
\begin{Schunk}
\begin{Sinput}
R> set.seed(10, kind = "L'Ecuyer-CMRG")
R> run_wei <- cure_rate_MC3(Surv(time, censoring) ~ age + kids + race,
+    data = marriage_dataset, nChains = nChains, mcmc_cycles = mcmc_cycles,
+    nCores = nCores, promotion_time = list(family = 'weibull'), 
+    verbose = FALSE)
\end{Sinput}
\begin{Soutput}
20 MCMC cycles required 1.21 secs. Expect a total run-time of: 904.89 secs. 
\end{Soutput}
\end{Schunk}
We can print some basic information and obtain a quick overview of the fitted models.
\begin{Schunk}
\begin{Sinput}
R> run_exp
\end{Sinput}
\begin{Soutput}
* Run information: 
      Fitted model: `exponential'
      BIC: 4119.714
      AIC: 4077.208
      MCMC cycles: 15000
      Number of parallel heated chains: 4
      Swap rates of adjacent chains: 
  Min. Median   Max. 
0.0012 0.0064 0.6277 

* Maximum A Posteriori (MAP) estimate of parameters 
                       MAP estimate
g_mcmc                    0.2221975
lambda_mcmc               2.3042290
a1_mcmc                   0.1412986
b0_mcmc [(Intercept)]     0.6332169
b1_mcmc [age]            -0.4825107
b2_mcmc [kidsyes]        -1.2673358
b3_mcmc [racehispanic]   -0.2779172
b4_mcmc [raceother]      -0.2361773
\end{Soutput}
\begin{Sinput}
R> run_wei
\end{Sinput}
\begin{Soutput}
* Run information: 
      Fitted model: `weibull'
      BIC: 4121.352
      AIC: 4073.533
      MCMC cycles: 15000
      Number of parallel heated chains: 4
      Swap rates of adjacent chains: 
  Min. Median   Max. 
  0.08   0.23   0.42 

* Maximum A Posteriori (MAP) estimate of parameters 
                       MAP estimate
g_mcmc                  -0.02222539
lambda_mcmc              4.81455675
a1_mcmc                  0.28933335
a2_mcmc                  0.60936921
b0_mcmc [(Intercept)]    0.79880702
b1_mcmc [age]           -0.47432163
b2_mcmc [kidsyes]       -1.31493290
b3_mcmc [racehispanic]  -0.23135416
b4_mcmc [raceother]     -0.23898757
\end{Soutput}
\end{Schunk}
Observe that both models produce similar point estimates for the common parameters of interest and particularly for the regression coefficients $\beta_j$, $j=1,\ldots,4$ which are denoted as \code{b0_mcmc, b1_mcmc, b2_mcmc, b3_mcmc, b4_mcmc} in the output above. The MAP estimate of $\gamma$ (denoted as \code{g_mcmc}) is also similar (0.22 versus -0.02). A somewhat larger deviation in MAP estimate is obtained for $\lambda$ (2.3 versus 4.81). The remaining parameter for the output of exponential model (\code{a1_mcmc}) refers to the rate parameter of the exponential distribution, as well as the rate (\code{a1_mcmc}) and shape parameter  (\code{a2_mcmc}) of the Weibull distribution in the output of the Weibull model. The BIC values (shown in the output above) are also returned using \proglang{R}'s generic function \fct{BIC}.
\begin{Schunk}
\begin{Sinput}
R> BIC(run_exp, run_wei)
\end{Sinput}
\begin{Soutput}
        df      BIC
run_exp  8 4119.714
run_wei  9 4121.352
\end{Soutput}
\end{Schunk}
The exponential model is  preferred and its full summary is shown below, after discarding the first 5000 iterations as burn-in period. 
\begin{Schunk}
\begin{Sinput}
R> burn <- mcmc_cycles/3
R> summary_exp <- summary(run_exp, fdr = 0.1, burn = burn, alpha0 = 0.1, 
+    quantiles = c(0.05,0.5,0.95))
R> summary_exp
\end{Sinput}
\begin{Soutput}
                       MAP_estimate        HPD_90%    5%   50%   95%
g_mcmc                         0.22 (-0.27,  0.80) -0.22  0.23  0.84
lambda_mcmc                    2.30   (1.96, 2.52)  1.97  2.23  2.52
a1_mcmc                        0.14   (0.10, 0.16)  0.10  0.13  0.16
b0_mcmc [(Intercept)]          0.63   (0.49, 0.93)  0.49  0.71  0.93
b1_mcmc [age]                 -0.48 (-0.56, -0.33) -0.56 -0.45 -0.34
b2_mcmc [kidsyes]             -1.27 (-1.46, -1.09) -1.46 -1.27 -1.09
b3_mcmc [racehispanic]        -0.28 (-0.48, -0.12) -0.48 -0.30 -0.12
b4_mcmc [raceother]           -0.24 (-0.45, -0.08) -0.45 -0.27 -0.09

Among 1018 censored observations, 231 items were identified as cured (FDR = 0.1). 
\end{Soutput}
\end{Schunk}

Note that we found 231 cured subjects when controlling the FDR at the \code{fdr = 0.1} level. The labels of these subjects can be returned by running e.g.~\code{which(summary_exp\$cured_at_given_FDR == "cured")}. The estimate of the marginal cure probability for each censored item is returned by calling \code{summary_exp$latent_cured_status}. Similarly, the argument \code{alpha0 = 0.1} produces $90\%$ Highest Posterior Density Intervals. The \code{quantiles} argument returns the corresponding sample quantiles of the retained MCMC sample. 

\begin{figure}[h]
\centering
\begin{Schunk}
\begin{Sinput}
R> par(mfrow = c(2,4), mar = c(4,3,2,2))
R> plot(run_exp, burn = burn, alpha0 = 0.1,
+    cex.axis = 2.5, cex.lab = 2.5, main = '', ylab = '')
\end{Sinput}
\end{Schunk}
\includegraphics{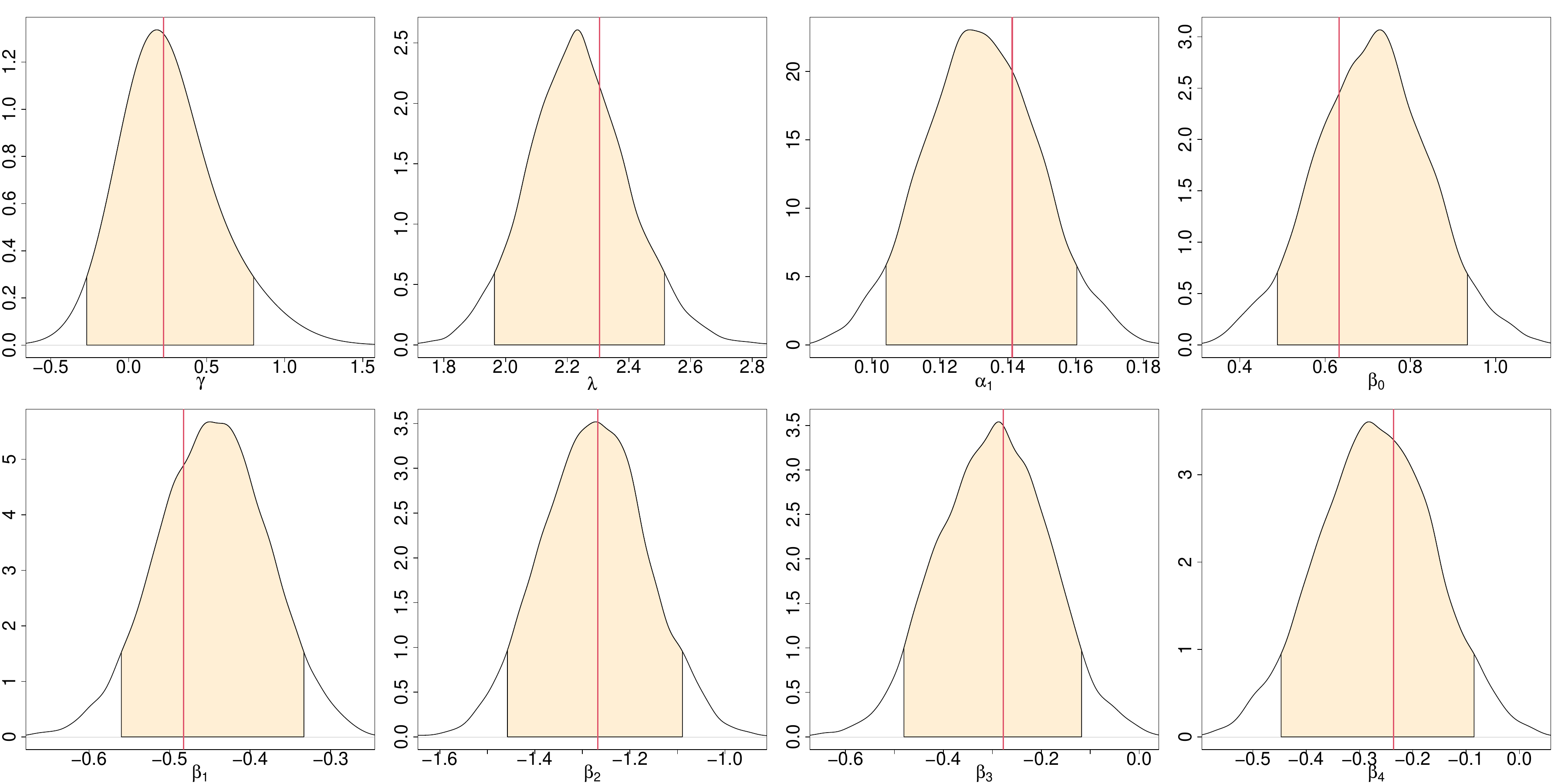}
\caption{\label{fig:plot1} Estimated marginal posterior distribution per parameter of the cure rate model with exponential distribution as promotion time. The shaded area corresponds to the $90\%$ Highest Posterior Density region. The vertical line corresponds to the Maximum A Posteriori estimate of the corresponding parameter arising from the joint posterior distribution.}
\end{figure}

The default \code{plot} method displays the estimated marginal posterior distribution for each parameter as shown in Figure \ref{fig:plot1}. 

\begin{figure}[h]
\centering
\begin{Schunk}
\begin{Sinput}
R> par(mfrow = c(2,2), mar = c(4,6,1,1))
R> plot(my_predictions1, what='survival',
+    ylim = c(0,1), cex.axis = 2.0, cex.lab = 2.5, draw_legend = FALSE)
R> plot(my_predictions2, what='survival',
+    ylim = c(0,1), cex.axis = 2.0, cex.lab = 2.5, draw_legend = FALSE)
R> plot(my_predictions1, what='cured_prob', 
+    ylim = c(0,1), cex.axis = 2.0, cex.lab = 2.5)
R> plot(my_predictions2, what='cured_prob', 
+    ylim = c(0,1), cex.axis = 2.0, cex.lab = 2.5)
\end{Sinput}
\end{Schunk}
\includegraphics{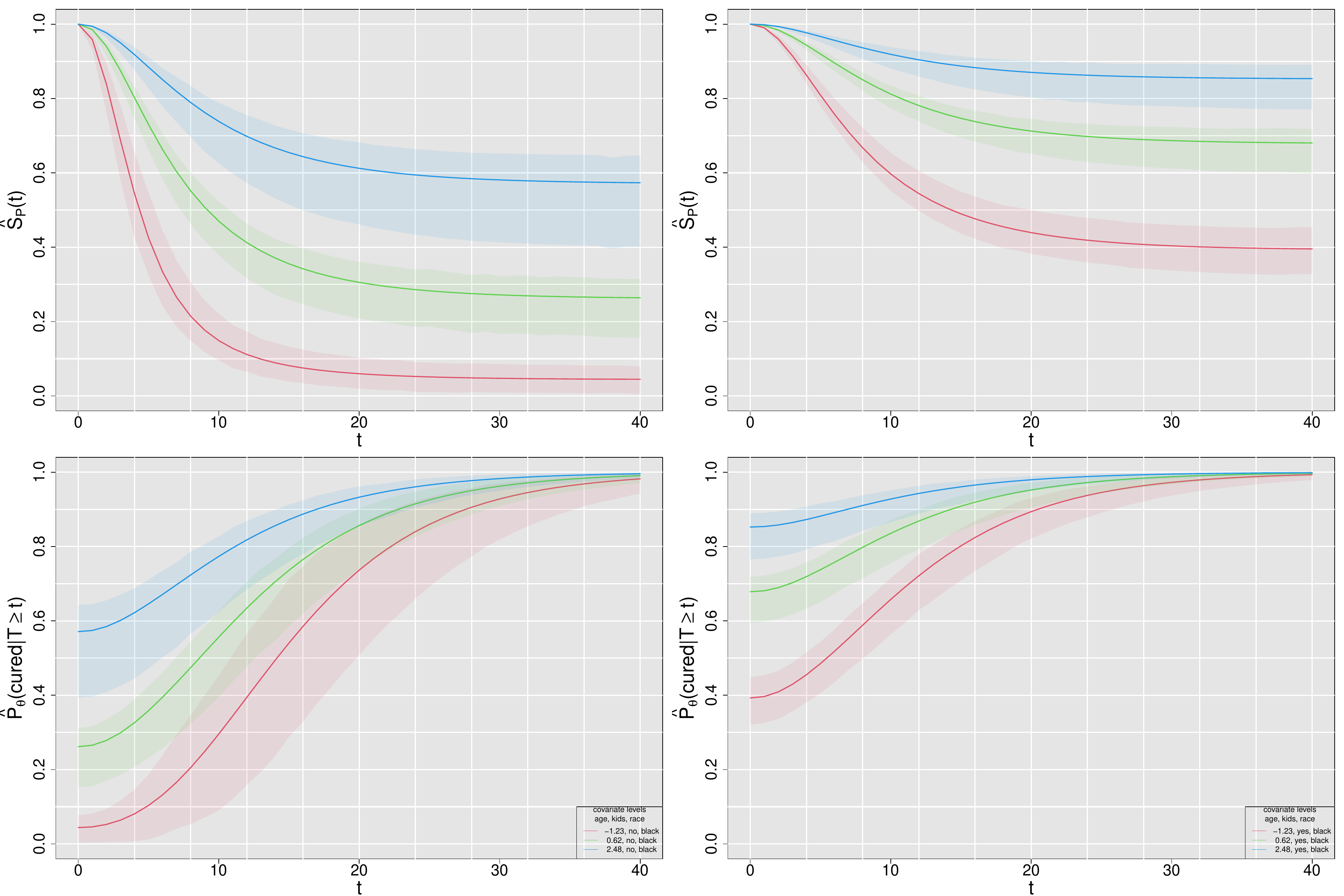}
\caption{\label{fig:plot2} Estimated survival function (top) and conditional cured probability (bottom) for various combinations of covariate levels. The left and right panels refer to the absence and presence of kids, respectively. The (scaled) age values correspond to 20 (-1.23), 30 (0.62) and 40 (2.48) years old. The highlighted area corresponds to (pointwise) $90\%$ credible intervals. }
\end{figure}

Next, let us retrieve model predictions for given covariate levels $\boldsymbol x$. We consider six different combinations of covariate levels in the \code{newdata} data frame, as shown below. 
\begin{Schunk}
\begin{Sinput}
R> age_mean <- as.numeric(attributes(marriage_dataset$age)[2])
R> age_sd <- as.numeric(attributes(marriage_dataset$age)[3])
R> x1 <- (20 - age_mean)/age_sd
R> x2 <- (30 - age_mean)/age_sd
R> x3 <- (40 - age_mean)/age_sd
R> covariate_levels1 <- data.frame(age = c(x1, x2, x3), kids = rep("no", 3), 
+    race = rep("black", 3))
R> covariate_levels2 <- data.frame(age = c(x1, x2, x3), kids = rep("yes", 3), 
+    race = rep("black", 3))
R> newdata <- rbind(covariate_levels1, covariate_levels2)
\end{Sinput}
\end{Schunk}
The three distinct values of the (standardized) age covariate correspond to 20, 30 and 40 years of age. Next, we call the \fct{predict} method which returns the estimates of survival probability $S_P(t)$, cumulative hazard rate $H_P(t) = -\log(S_P(t))$, hazard rate $h_P(t) = f_P(t)/S_P(t)$ and the conditional cured probability $P(I = 0|T\geqslant t)$, for $t = 10, 20$ years. 
\begin{Schunk}
\begin{Sinput}
R> my_predictions <- predict(run_exp, newdata = newdata, 
+    tau_values = c(10, 20), alpha0 = 0.1)
R> my_predictions
\end{Sinput}
\begin{Soutput}
$`t = 10`
     age kids  race S_p[t]     S_p[t]_90% H_p[t]     H_p[t]_90%
1 -1.231   no black  0.149 (0.096, 0.222)  1.904 (1.453, 2.269)
2  0.624   no black  0.470 (0.379, 0.523)  0.754 (0.643, 0.966)
3  2.479   no black  0.738 (0.625, 0.789)  0.303 (0.233, 0.465)
4 -1.231  yes black  0.597 (0.552, 0.654)  0.516 (0.422, 0.591)
5  0.624  yes black  0.812 (0.775, 0.834)  0.208 (0.181, 0.256)
6  2.479  yes black  0.919 (0.880, 0.938)  0.085 (0.063, 0.126)
  h_p[t]     h_p[t]_90% P[cured|T > t] P[cured|T > t]_90%
1  0.163 (0.104, 0.224)          0.296     (0.090, 0.453)
2  0.073 (0.059, 0.095)          0.556     (0.395, 0.625)
3  0.031 (0.023, 0.048)          0.773     (0.624, 0.827)
4  0.051 (0.042, 0.059)          0.658     (0.564, 0.716)
5  0.021 (0.018, 0.027)          0.835     (0.755, 0.870)
6  0.009 (0.006, 0.014)          0.928     (0.865, 0.951)

$`t = 20`
     age kids  race S_p[t]     S_p[t]_90% H_p[t]     H_p[t]_90%
1 -1.231   no black  0.060 (0.019, 0.102)  2.817 (2.121, 3.600)
2  0.624   no black  0.305 (0.208, 0.361)  1.186 (1.003, 1.547)
3  2.479   no black  0.612 (0.462, 0.682)  0.491 (0.379, 0.769)
4 -1.231  yes black  0.439 (0.383, 0.500)  0.823 (0.689, 0.954)
5  0.624  yes black  0.713 (0.652, 0.745)  0.339 (0.294, 0.428)
6  2.479  yes black  0.870 (0.803, 0.901)  0.139 (0.102, 0.217)
  h_p[t]     h_p[t]_90% P[cured|T > t] P[cured|T > t]_90%
1  0.043 (0.021, 0.082)          0.737     (0.508, 0.857)
2  0.021 (0.016, 0.035)          0.857     (0.739, 0.901)
3  0.010 (0.007, 0.018)          0.933     (0.855, 0.962)
4  0.015 (0.012, 0.022)          0.894     (0.821, 0.930)
5  0.007 (0.005, 0.011)          0.952     (0.905, 0.972)
6  0.003 (0.002, 0.006)          0.980     (0.950, 0.990)
\end{Soutput}
\end{Schunk}
In the code above, the \code{alpha0} argument specifies the credibility level of  the corresponding Highest Posterior Density intervals of each quantity. The user can also pass the \fct{predict} output to the \fct{plot} method in order to effectively visualize predictions, as shown below. For this purpose it is better to use a more detailed sequence of $t$ values in the \code{tau_values} argument. 
\begin{Schunk}
\begin{Sinput}
R> tau_values <- seq(0, 40, by = 1)
R> my_predictions1 <- predict(run_exp, newdata = covariate_levels1, 
+    tau_values = tau_values, alpha0 = 0.1)
R> my_predictions2 <- predict(run_exp, newdata = covariate_levels2, 
+    tau_values = tau_values, alpha0 = 0.1)
\end{Sinput}
\end{Schunk}
Figure \ref{fig:plot2} illustrates the survival function as well as the estimated cured probability, conditional on the event that the subject has survived until time $t$, for the aforementioned covariate levels, after calling the \code{plot} command as  shown below. 

\begin{figure}[ht]
\centering
\begin{Schunk}
\begin{Sinput}
R> par(mfrow = c(2,5), mar = c(4,5,2,2))
R> plot(run_wei, burn = burn, 
+    cex.axis = 2.5, cex.lab = 2.5, main = '', ylab = '')
R> thin_sequence <- seq(burn, mcmc_cycles, by = 10)
R> plot(run_wei$mcmc_sample[thin_sequence, 'g_mcmc'], 
+    run_wei$log_posterior[thin_sequence], 
+    xlab = bquote(gamma), ylab = 'log-posterior density',
+    cex.axis = 2.5, cex.lab = 2.5, col = 'blue')
\end{Sinput}
\end{Schunk}
\includegraphics{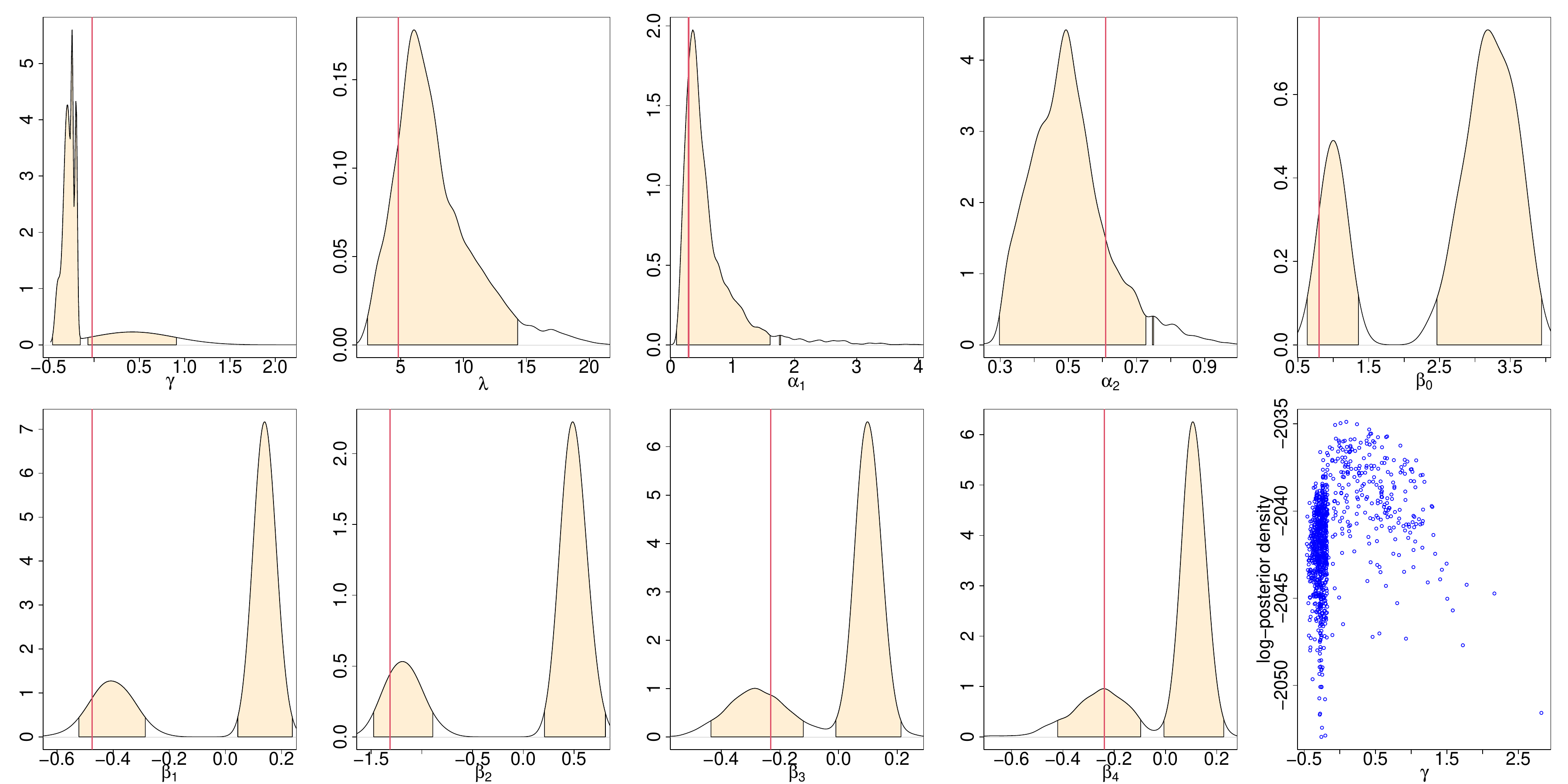}
\caption{\label{fig:plot1wei} Estimated marginal posterior distribution per parameter of the cure rate model with Weibull distribution as promotion time. The shaded area corresponds to the $95\%$ Highest Posterior Density region. The vertical line corresponds to the Maximum A Posteriori estimate of the corresponding parameter arising from the joint posterior distribution. The last panel displays (a thinned subset of) the sampled values of $\gamma$ versus the corresponding values of the logarithm of the posterior distribution (up to a normalizing constant). }
\end{figure}

Next we have a closer look at the output of the Weibull model in Section \ref{sec:illustrations}. As shown in Figure \ref{fig:plot1wei}, the aforementioned multimodality in the posterior draws is evident and it is particularly notable for $\gamma$, $\beta_0$, $\beta_1$, $\beta_2$, $\beta_3$ and $\beta_4$. In order to shed further light into this aspect, the last panel of Figure \ref{fig:plot1wei} displays (a thinned subset of) the sampled values of $\gamma$ versus the corresponding values of the logarithm of the posterior distribution $\pi(\boldsymbol\theta|\boldsymbol y, \boldsymbol x)$ (up to a normalizing constant). Clearly, the sampled values of $\gamma$ form  distinct modes: the positive draws (around $0.5$) come from the main mode where the maximum values of the log-posterior are attained, while the negative draws  correspond to slightly smaller values of the log-posterior density function. 

\begin{figure}[h]
\centering
\begin{Schunk}
\begin{Sinput}
R> par(mfrow = c(1,2))
R> plot(run_exp, what = 'residuals', main = 'Exponential', 
+    ylab = 'Kaplan-Meier cumulative hazard')
R> plot(run_wei, what = 'residuals', main = 'Weibull', 
+    ylab = 'Kaplan-Meier cumulative hazard')
\end{Sinput}
\end{Schunk}
\includegraphics{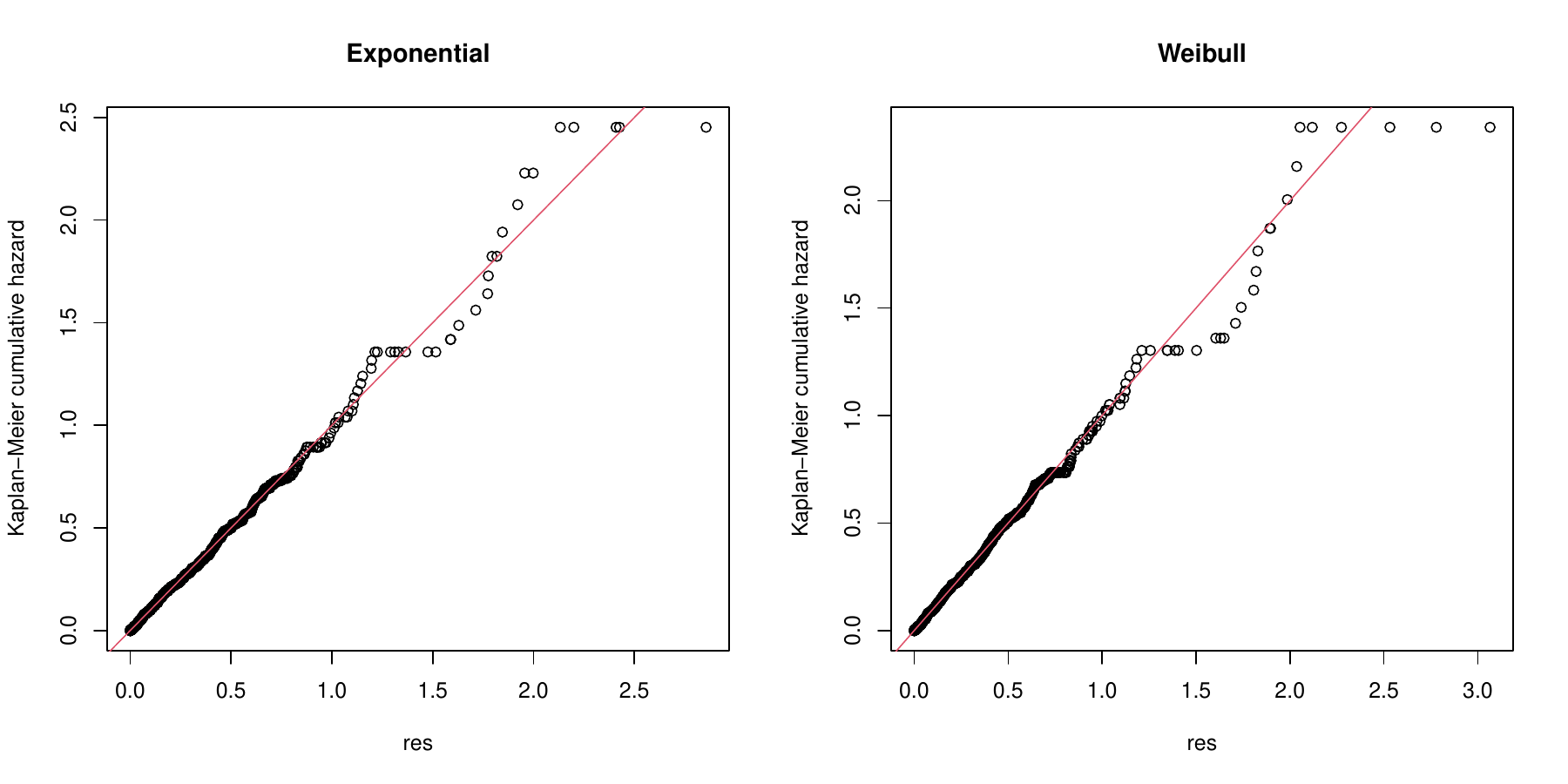}
\caption{\label{fig:residual_plot} Plot of the Cox-Snell residuals versus the estimated cumulative hazard as obtained from the Kaplan-Meier estimator. }
\end{figure}

Although the literature on model diagnostics for cure rate modeling is not yet extensive, we suggest using Cox-Snell residuals (\citealp{cox1968general}) to assess the overall fit of 
model \eqref{eq0001}. The properties of these residuals under the mixture cure rate model have been studied by, for example, \cite{peng2017residual} and \cite{scolas2018diagnostic}. If 
the time-to-event variable \( T \) indeed follows the survival function given in \eqref{eq0001}, then
\begin{align*}
P(-\log(S_P(T))<t)=P(S_P(T)>e^{-t})=\begin{cases}
1, & t\geq  -\log(p_0) \\
e^{-t}, & t<-\log(p_0)
\end{cases},
\end{align*}
where $p_0$ is the cure rate inferred from model \eqref{eq0001}. 
This indicates that the Cox-Snell residuals \( r_{CS}(y_i) = -\log(S_P(y_i)) \), for \( i = 1, \ldots, n \), should behave like a censored sample from an exponential distribution with a mean equal to one, for \( t \in [0, -\log(p_0)) \), assuming the model is correct 
(it is necessary to mention that \( r_{CS}(y_i) \in [0, -\log(p_0)) \), for every $y_i$). Consequently, similar to classical survival models or mixture cure model, 
a plot of \( r_{CS}(y_i) \) against their estimated cumulative hazard, as obtained, for example, from the Kaplan-Meier estimator, should ideally show points lying close to the 45-degree line (recall that the cumulative hazard function of an exponential distribution with mean equal to one, is the identity function). These values are plotted in Figure \ref{fig:residual_plot} for the exponential and Weibull models and we do observe that the points are close to the 45-degree line as expected. 

We close this section by mentioning that we have also fit all remaining choices for the promotion time distribution in this dataset (namely, the gamma, mixture of gamma with 2 and 3 components, Gompertz, log-logistic, Lomax and Dagum models) but the exponential model was ranked first according to the BIC (results not shown). 


\section{Summary and discussion} \label{sec:summary}

The contributed package can be used to estimate cure rate models under a Bayesian setup, building upon the methodology introduced in \cite{papastamoulis2023bayesian}. The underlying family of cure rate models was originally introduced in \cite{milienos2022reparameterization} and includes various models (such as the promotion time, the negative binomial and the mixture cure rate model) as specific cases. Naturally, the likelihood and posterior surface may be multimodal in order to accomodate all these special cases and this burdens the estimation procedure both under frequentist as well as Bayesian perspectives. The proposed methodology provides a practical means of performing robust Bayesian inference using a tailored Metropolis-Coupled MCMC sampler. 

We recommend to use our method by calling the main function (\code{cure_rate_MC3}) with at least 15000 MCMC cycles (\code{mcmc_cycles}) and a minimum of $4$ heated chains (\code{nChains}). According to our experimentation with real datasets, we suggest trying at least  the Weibull model, however the user can fit all available choices and select one according to information criteria, such as the BIC. We also suggest to scale all continuous covariates so their sample mean and sample variance are equal to 0 and 1, respectively. In Unix-like systems we recommend to enable parallelization, by using at least 4 cores (\code{nCores}), but it is preferable to retain just one core in Windows (see the ``Computational details'' section).

We did not address the issue of variable selection. Of course, one can compare various models using information criteria such as the BIC. However, we plan to explore this issue in the future using Bayesian variable selection techniques, such as stochastic search variable selection (see e.g.~\cite{george1995stochastic, dellaportas2002bayesian}) or adopting shrinkage priors \citep{10.1093/acprof:oso/9780199694587.003.0017}. 

All results in Section \ref{sec:illustrations} were obtained using a Linux workstation with the following specifications: Operating System: Ubuntu 24.04.2 LTS, 64-bit, Processor: Intel Core i9-11900 @ $2.50$GHz $\times$ 16. The following versions of linear algebra libraries were used:  libblas.so.3.12.0 (BLAS) and liblapack.so.3.12.0 (LAPACK). The job-script ran using a single core. Note that the   number of cores is important to reproduce results (under the same seed), but the results are not reproducible in case of different operating systems and/or other versions of linear algebra libraries (see also the Computational details section).


\section*{Computational details}

Our implementation when considering a large number of heated chains can take advantage of parallel processing in certain cases. In brief, the \code{nChains} heated chains are distributed among the \code{nCores} available cores. The \code{nCores} workers are stopped at the end of each MCMC cycle in order to perform the swap move between adjacent chains and start again. This procedure is repeated for a total of \code{mcmc\_cycles}. For this purpose, the libraries \pkg{foreach} \citep{foreach} and \pkg{doParallel} \citep{doparallel} were used. However, the practical gain of parallel computations depends on the Operating System, as detailed below.

\begin{figure}[ht]
\centering
\begin{Schunk}
\begin{Sinput}
R> library(ggplot2)
R> df <- read.csv('run_times.csv')
R> df2 <- aggregate(Time ~ Cores + OS, data = df, FUN= "mean" )
R> ggplot(df2,  aes(x = Cores, y = Time, group = OS, color = OS)) +
+      geom_line() + geom_point() +  scale_y_log10()
\end{Sinput}
\end{Schunk}
\includegraphics{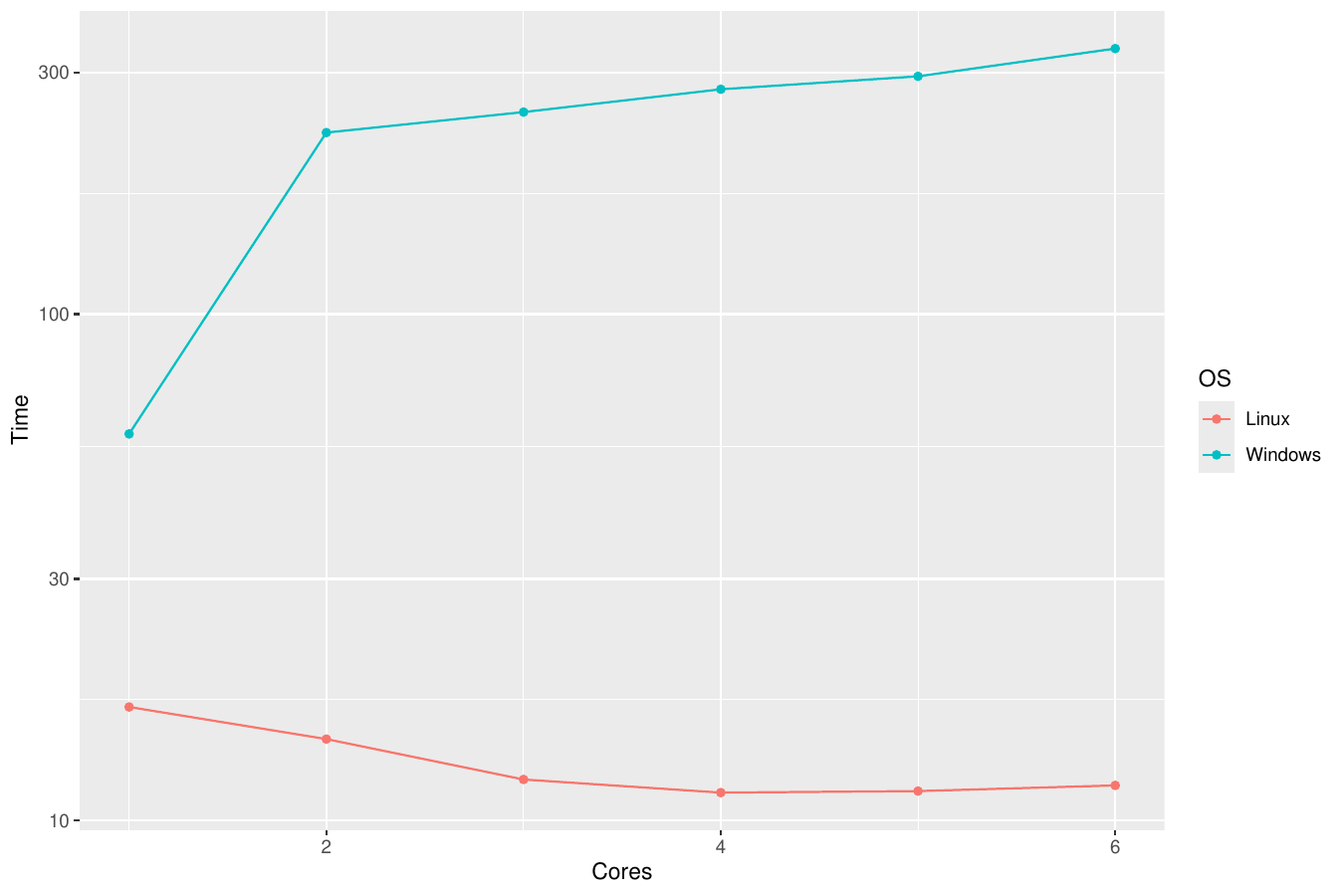}
\caption{\label{fig:time2} The time (in seconds) required to run \code{mcmc\_cycles = 100} MCMC cycles with \code{nChains = 12} heated chains using various number of cores (\code{nCores}) on a Linux/Ubuntu versus a Windows workstation. Details: sample size $n = 1500$ observations, $p=5$ covariates (including constant term). Each point in the graph corresponds to the average elapsed time arising from four distinct runs. The $y$-axis is on $\log_{10}$ scale. }
\end{figure}

Figure \ref{fig:time2} compares the elapsed run-time required to run 12 heated chains for a total of 100 MCMC cycles as a function of the number of cores. We conclude that parallelization reduces significantly the run-time when using up to 3 or 4 cores and a Linux workstation. However, this is not the case for Windows: observe that the run-time is increased dramatically when distributing the computation into parallel chains, therefore we recommend to disable parallelization in Windows (that is, using \code{nCores = 1}). The results were obtained using two workstations with the following specifications: 
\begin{enumerate}
\item Linux workstation details: OS: Ubuntu 24.04.2 LTS, 64-bit, Processor: Intel Core i9-11900 @ $2.50$GHz $\times$ 16. 
\item Windows workstation details: OS: Windows 11 Home 64-bit, Processor: Intel Core i7-9750 @ $2.60$GHz $\times$ 12.
\end{enumerate}

Recall that in  \proglang{R}, parallel computation can be achieved using different types of clusters, that is ``PSOCK'' (Socket) clusters and ``FORK'' clusters. PSOCK clusters are available in both Windows and Unix, however they tend to be much slower than FORK clusters which are only available in Unix-like systems (e.g.~Linux, MacOS). The necessity of using sockets and serializing data for inter-process communication on Windows (PSOCK) introduces additional overhead, making the process slower and less efficient than FORK clusters, which are ideal for parallel computing in Unix-like systems.

We have also used the \pkg{Rcpp}   \citep{Rcpp1, Rcpp2} and \pkg{RcppArmadillo} \citep{RcppArmadillo} packages in order to compute the log-likelihood in Equation \eqref{eq:cll}. The gradient vector in Equation \eqref{eq:proposal} has been computed numerically using the \pkg{calculus} package \citep{calculus}. Highest posterior density intervals have been computed using the \pkg{HDInterval} package \citep{HDI}.

We should mention here that the results of Section \ref{sec:illustrations} are not reproducible when at least one of the following conditions is not met: 
\begin{enumerate}
\item same number of cores (that is, \code{nCores = 1})
\item same Operating System (that is, Ubuntu 24.04.2).
\end{enumerate}
Note that the option \code{kind = "L'Ecuyer-CMRG"} \citep{l1999good} (used in our calls to the \code{set.seed} command) is suggested when using  (\code{nCores} $> 1$) for reproducible random number generation. We have also run our code on both Windows and Mac workstations but we weren't able to reproduce the results, despite fixing the seed and the number of cores. The explanation for this behaviour is that when using the \pkg{Rcpp} library in \proglang{R}, differences in reproducibility between different operating systems can occur due to factors as floating-point arithmetic, different compilers and different library versions (such as BLAS, LAPACK, or other numeric libraries). Unfortunately, these factors are not easy to control. However, we mention that the differences we obtained are not worth mentioning and all conclusions remain valid, since the MCMC sampler has achieved convergence.

\section*{Acknowledgments}

Panagiotis Papastamoulis received funding from the Research Center of Athens University of Economics and Business. The authors would like to thank an anonymous Editor and two reviewers of the JSS whose comments substantially improved the content of the package and the presentation of our findings. 


\bibliography{refs}

\newpage

\begin{appendix}

\section{Parameterizations of distributions} \label{app:technical}

The probability density function of the distributions used in this paper are parameterized as follows. 
\begin{itemize}
\item Exponential distribution with rate parameter $\alpha_1 > 0$
\[
f(y) = \alpha_1 e^{-\alpha_1 y},\quad y > 0.
\]
\item Gompertz distribution with shape $\alpha_1>0$ and rate $\alpha_2>0$
\[
f(y) = \alpha_2 e^{\alpha_1 y}e^{-\frac{\alpha_2}{\alpha_1} \{e^{\alpha_1 y} - 1\}},\quad y > 0 
\]
as implemented in the \pkg{flexsurv} \citep{flexsurv} package.
\item log-logistic distribution with shape parameter $\alpha_1 > 0$ and scale parameter $\alpha_2 > 0$
\[
 f(y) = \frac{\alpha_1 y^{\alpha_1-1}}{\alpha_2^{\alpha_1}}  \left\{1 + \left(\frac{y}{\alpha_2}\right)^{\alpha_1}\right\}^{-2},\quad y > 0
\]
as implemented in the \pkg{flexsurv} \citep{flexsurv} package.
\item Weibull distribution with rate $\alpha_1 > 0$ and shape $\alpha_2 > 0$
\[
f(y) = \alpha_1\alpha_2 \alpha_1^{\alpha_2 - 1} y^{\alpha_2-1}e^{-(\alpha_1 y)^{\alpha_2}},\quad y > 0.
\]
\item gamma distribution $\mathcal{G}(\alpha_1,\alpha_2)$ with shape $\alpha_1 > 0$ and rate $\alpha_2 > 0$
\[f_(x)=\frac{\alpha_2^{\alpha_1}}{\Gamma(\alpha_1)}y^{\alpha_1-1}\exp\{-\alpha_2 y\}, \quad y>0.\]
\item Inverse gamma $\mathcal IG(\alpha_1,\alpha_2)$ with shape $\alpha_1 > 0$ and scale $\alpha_2 > 0$
\[
 f(y) = \frac{\alpha_2^{\alpha_1}}{\Gamma(\alpha_1)}\frac{1}{y^{\alpha_1+1}}e^{-\frac{\alpha_2}{y}},\quad y > 0.
\]
\item Lomax distribution with shape parameter $\alpha_1 > 0$ and scale parameter $\alpha_2 > 0$
\[
f(y) = \frac{\alpha_1}{[\alpha_2 (1 + y/\alpha_2)^{1+\alpha_1}]}, \quad y > 0
\]
as implemented in the \pkg{VGAM} package \citep{vgam}.
\item Dagum distribution with scale parameter $\alpha_1 > 0$ and shape parameters $\alpha_2 > 0$ and $\alpha_3 > 0$
\[
f(y) = \frac{\alpha_2 \alpha_3}{\alpha_1} \left(\frac{y}{\alpha_1}\right)^{\alpha_2 \alpha_3-1} \left[1 + \left(\frac{y}{\alpha_1}\right)^{\alpha_2}\right]^{-(\alpha_3+1)},\quad y > 0
\]
as implemented in the \pkg{VGAM} package \citep{vgam}.
\end{itemize}

\section{User-defined distributions}\label{app:user}

In this section we illustrate how the user can fit custom families of (univariate) distributions, as well as finite mixtures of these families, for describing the promotion time. The only restriction is that the custom-defined families should be parameterized in such a way so that all parameters belong to the set $(0,\infty)$. 

\begin{figure}[h]
\centering
\begin{Schunk}
\begin{Sinput}
R> library(survival)
R> km_fit <- survfit(Surv(time, censoring) ~ 1, data = sim_mix_data)
R> par(mfrow = c(1,2), mar = c(4, 4, 1,1))
R> plot(km_fit, conf.int = FALSE, mark.time = TRUE, xlab = 'time', 
+    cex = 0.5, pch = 4)
R> plot(km_fit, conf.int = FALSE, mark.time = TRUE, xlab = 'time', 
+    xmax = 10, cex = 0.5, pch = 4)
\end{Sinput}
\end{Schunk}
\includegraphics{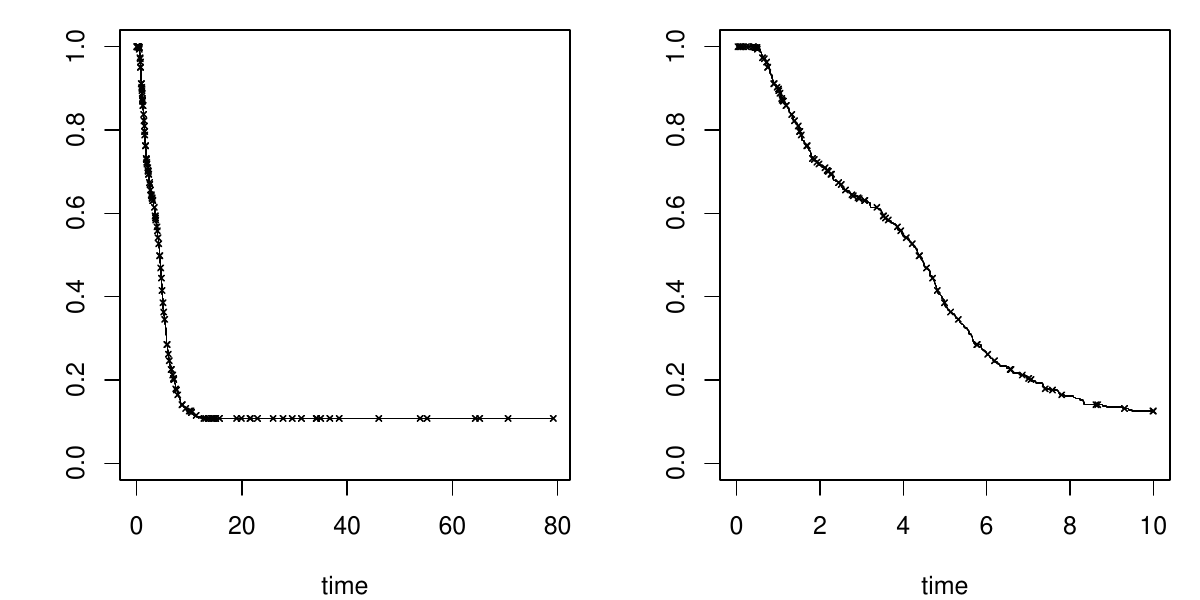}
\caption{\label{fig:hist} Kaplan-Meier survival curve for the synthetic dataset. The right panel is a zoomed version of the left panel.}
\end{figure}

We will consider a synthetic dataset which is part of the \pkg{bayesCureRateModel} package.
\begin{Schunk}
\begin{Sinput}
R> data(sim_mix_data)
R> str(sim_mix_data, strict.width = 'cut')
\end{Sinput}
\begin{Soutput}
'data.frame':	500 obs. of  5 variables:
 $ time       : num  0.691 6.157 2.914 2.796 3.147 ...
 $ censoring  : num  1 1 0 0 1 1 1 1 1 0 ...
 $ x1         : num  0.9466 0.7431 0.0508 0.9804 0.2178 ...
 $ x2         : Factor w/ 3 levels "0","1","2": 1 3 3 3 2 1 3 1 1 3 ..
 $ true_status: Factor w/ 2 levels "cured","susceptible": 2 2 1 2 2 ..
\end{Soutput}
\begin{Sinput}
R> table(sim_mix_data$true_status)
\end{Sinput}
\begin{Soutput}
      cured susceptible 
         59         441 
\end{Soutput}
\end{Schunk}
There are two covariates in this dataset with column names \code{x1} and \code{x2}. The column \code{true_status} contains the true (latent) status of each observation. There are 59 cured subjects in total. At first, we can inspect the observed times (\code{time}), as shown in Figure \ref{fig:hist}.
Suppose that the user wishes to fit the proposed model using a mixture of two distributions in order to describe promotion time. Let us pick the family of log-normal distributions for each component. In reality, the synthetic dataset has been generated by model \eqref{eq0001} considering a mixture of two gamma distributions for describing the promotion time, under an exponential censoring scheme. We will fit two models in total: (a) a simple log-normal model and (b) a mixture of two log-normal distributions.

At first, we define a function which returns the logarithm of the probability density function and cumulative density function of the log-normal distribution. Let us recall that the probability density function of the log-normal distribution is typically defined as 
\[
f(y; \mu, \sigma) = \frac{1}{y \sigma \sqrt{2\pi}} \exp\left( -\frac{(\log y - \mu)^2}{2 \sigma^2} \right), \quad y > 0,
\] 
where $\mu\in(-\infty,\infty)$ and $\sigma > 0$. In our package, each parameter of a user-defined function should lie on the set $(0, \infty)$, so we have to reparameterize the previous density as follows:
\begin{equation}
\label{eq:ln}
f(y; a_1, a_2) = \frac{1}{y a_2 \sqrt{2\pi}} \exp\left( -\frac{(\log y - \log a_1)^2}{2 a_2^2} \right), \quad y > 0,
\end{equation}
that is, $a_1 = e^\mu$ and $a_2 = \sigma$. We will use the notation $\mathcal{LN}(a_1,a_2)$ to refer to the family of log-normal distributions in \eqref{eq:ln}, $a_1 > 0$ and $a_2 > 0$. Next, we define a function that computes the $\log f(y;a_1,a_2)$ and $\log \int_{0}^{y} f(t;a_1,a_2)\mathrm{d}t$, as follows.
\begin{Schunk}
\begin{Sinput}
R> user_promotion_time <- function(y, a){
+    log_f <- -0.5*log(2*pi) - log(y) - log(a[2]) - 
+    ((log(y) - log(a[1]))^2)/(2 * a[2]^2)
+    log_F <- pnorm((log(y) - log(a[1]))/a[2], log.p = TRUE)
+    result <- vector('list', length = 2)
+    names(result) <- c('log_f', 'log_F')
+    result[["log_f"]] = log_f
+    result[["log_F"]] = log_F
+    return(result)
+  }
\end{Sinput}
\end{Schunk}
As seen in the code snippet above,  the user-defined function should always accept two arguments \code{y} (corresponding to the data) and \code{a} (vector of positive parameters). In addition, it should always return a list with named arguments \code{log_f} and \code{log_F} corresponding to the logarithm of the probability density function and cumulative density function, respectively. Now, we can fit a simple log-normal model, as follows. 
\begin{Schunk}
\begin{Sinput}
R> promotion_time <- list(family = "user", 
+  		define = user_promotion_time,
+  		prior_parameters = matrix(rep(c(2.1, 1.1), 2), 
+  		byrow = TRUE, 2, 2), prop_scale = c(0.1, 0.1)
+  	)
\end{Sinput}
\end{Schunk}
As shown above, the user has to define a list (\code{promotion_time}) which contains the following entries:
\begin{itemize}
\item \code{family = "user"} which means that a user-defined family of distributions is going to be fitted.
\item \code{define} the function which accepts as input the data (\code{y}) and a vector (\code{a}) of positive parameters and returns the logarithm of the probability density (\code{log_f}) function and cumulative density function (\code{log_F}) in the form of a list.
\item \code{prior_parameters} is a matrix containing as many rows as the length of the parameters (here it is equal to two), and the columns contain the values of the prior distributions, that is, $\mathcal{IG}(2.1, 1.1)$ for both parameters.
\item \code{prop_scale} contains the scale of the random-walk proposal in the Metropolis-Hastings step of the sampler. 
\end{itemize}
\end{appendix}
After this step, the user can call the \fct{cure\_rate\_MC3} as usual. For illustration purposes we are going to use 1000 MCMC cycles. 
\begin{Schunk}
\begin{Sinput}
R> set.seed(1, kind = "L'Ecuyer-CMRG")
R> run_ln <- cure_rate_MC3(survival::Surv(time, censoring) ~ x1 + x2, 
+    data = sim_mix_data, mcmc_cycles = 1000, promotion_time = promotion_time, 
+    nChains = 4, nCores = 1, verbose = FALSE)
\end{Sinput}
\begin{Soutput}
20 MCMC cycles required 0.54 secs. Expect a total run-time of: 27.09 secs. 
\end{Soutput}
\end{Schunk}
Next, we exemplify how to fit a model where the promotion time follows a finite mixture of $K$ log-normal distributions of the form $\sum_{k=1}^{K}w_k\mathcal{LN}(y;a_{1k},a_{2k})$, where $w_j > 0$ and $\sum_{k=1}^{K}w_k=1$, while $a_{ik}>0$ for $i=1,2$ and $k=1,\ldots,K$, for a given number of components $K > 1$. We will assume a two component mixture, that is, $K=2$.
\begin{Schunk}
\begin{Sinput}
R> K <- 2
R> n_f <- 2
R> prior_parameters <- array(data = NA, dim = c(n_f,2,K))
R> for(k in 1:K){
+  	prior_parameters[,,k] = matrix(rep(c(2.1, 1.1), n_f), 
+  	byrow = TRUE, n_f, 2)}
\end{Sinput}
\end{Schunk}
In the code snippet above, \code{K} defines the number of components of the finite mixture model, \code{n_f} denotes the number of component-specific parameters, that is, 2 in our case. The object \code{prior_parameters} is  a \code{n_f}$\times$\code{2}$\times$\code{K}-dimensional array, containing the values of the inverse gamma prior distributions for each parameter of the mixture components. In this case, we are assuming a-priori that $a_{ik}\sim\mathcal{IG}(2.1, 1.1)$, independent for $i=1,2$ and $k = 1,\ldots,K$. Next we have to pass the remaining ingredients of the model in the \code{promotion_time} argument of the main function, as follows
\begin{Schunk}
\begin{Sinput}
R> promotion_time <- list(family = 'user_mixture', 
+    define = user_promotion_time, prior_parameters = prior_parameters, 
+    prop_scale =  rep(0.1, K*n_f + K - 1), K = K, 
+    dirichlet_concentration_parameter = 1)
\end{Sinput}
\end{Schunk}
Notice that the argument \code{family} now is set to \code{"user_mixture"} while the \code{define} argument is set to the same input as previously. This will instruct the main function of our package to fit a mixture of log-normal densities. Finally, the \code{dirichlet_concentration_parameter} specifies the concentration parameter of the underlying Dirichlet prior distribution of the mixing proportions. After this step, the user can call the \fct{cure\_rate\_MC3} as usual. For illustration purposes we are going to use 1000 MCMC cycles. 
\begin{Schunk}
\begin{Sinput}
R> set.seed(1, kind = "L'Ecuyer-CMRG")
R> run_ln_mix <- cure_rate_MC3(survival::Surv(time, censoring) ~ x1 + x2, 
+    data = sim_mix_data, mcmc_cycles = 1000, promotion_time = promotion_time, 
+    nChains = 4, nCores = 1, verbose = FALSE)
\end{Sinput}
\begin{Soutput}
20 MCMC cycles required 5.81 secs. Expect a total run-time of: 290.4 secs. 
\end{Soutput}
\end{Schunk}

\begin{figure}[ht]
\centering
\includegraphics{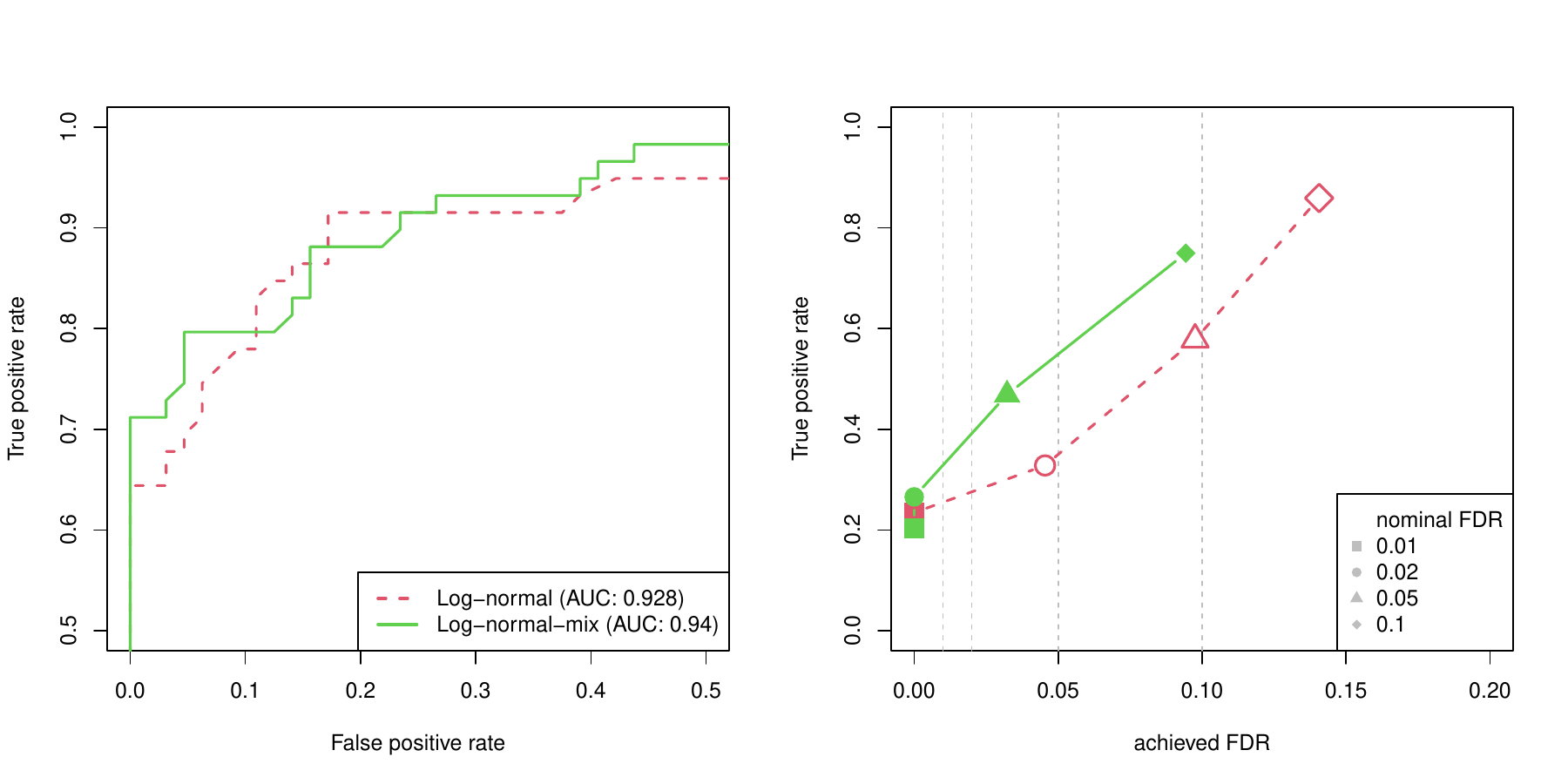}
\caption{\label{fig:classification} ROC curve (left) and power versus achieved FDR diagram (right) for the log-normal and mixture of two log-normals models. }
\end{figure}

Now, we can compare the two models using the BIC, as follows
\begin{Schunk}
\begin{Sinput}
R> BIC(run_ln, run_ln_mix)
\end{Sinput}
\begin{Soutput}
           df      BIC
run_ln      8 1603.316
run_ln_mix 11 1491.033
\end{Soutput}
\end{Schunk}
and we conclude that the mixture model should be preferred, as expected. Next we can evaluate the ability of the two models to correctly identify items as cured or not, since we do have the ground-truth status of each item in our simulated dataset. For this purpose we will use a ROC curve as well as a plot of the achieved FDR versus the True Positive Rate (see, e.g.~\citealp{Soneson2016}), for a series of nominal FDR levels. This can be done using the \pkg{ROCR} package \citep{rocr} and the following code.
\begin{Schunk}
\begin{Sinput}
R> library(ROCR)
R> ss_ln <- summary(run_ln, burn = 300, verbose = FALSE)
R> ss_ln_mix <- summary(run_ln_mix, burn = 300, verbose = FALSE)	
R> latent_cured_status_ln <- ss_ln$latent_cured_status
R> latent_cured_status_ln_mix <- ss_ln_mix$latent_cured_status
R> labels <- sim_mix_data$true_status[sim_mix_data$censoring == 0]
R> labels <- factor(labels, levels = c('susceptible', 'cured'), 
+    ordered = TRUE)
R> pred_ln <- prediction(latent_cured_status_ln, labels)
R> pred_ln_mix <- prediction(latent_cured_status_ln_mix, labels)
R> perf_ln <- performance(pred_ln, "tpr", "fpr")
R> perf_ln_mix <- performance(pred_ln_mix, "tpr", "fpr")
R> myCut = c(1,2, 5,10)/100
R> true_latent_status <- as.numeric(labels) - 1
R> fdr_tpr_ln <- compute_fdr_tpr(true_latent_status, latent_cured_status_ln, 
+    myCut = myCut)
R> fdr_tpr_ln_mix <- compute_fdr_tpr(true_latent_status, 
+    latent_cured_status_ln_mix, myCut = myCut)
\end{Sinput}
\end{Schunk}
Figure \ref{fig:classification} displays the resulting ROC curve and power versus achieved diagrams. On the latter graph, a coloured symbol indicates the corresponding FDR is controlled at the nominal. We conclude that the mixture model is able to control the FDR rate within the desired limits, something that is not true for the simple log-normal model when the nominal FDR is equal to $0.05$ or $0.10$. At the same time, the mixture model exhibits high discriminative power. 

Finally, we close this section by mentioning that the MCMC output of the component specific parameters is not identifiable due to the label switching problem \citep{redner1984mixture} of finite mixture models and we note that there is a variety of available methods for dealing with this issue \citep{papastamoulis2010artificial, labelSwitching}. However, the main inferential tasks here are unaffected by the labeling of the mixture components (such as estimation of the survival function, cure rate and identification of items as cured or not). 

\section{Comparison against alternative approaches}
\label{sec:comparison}

\subsection{Comparison with STAN}
\label{sec:stan}
\begin{figure}[ht]
\centering
\includegraphics{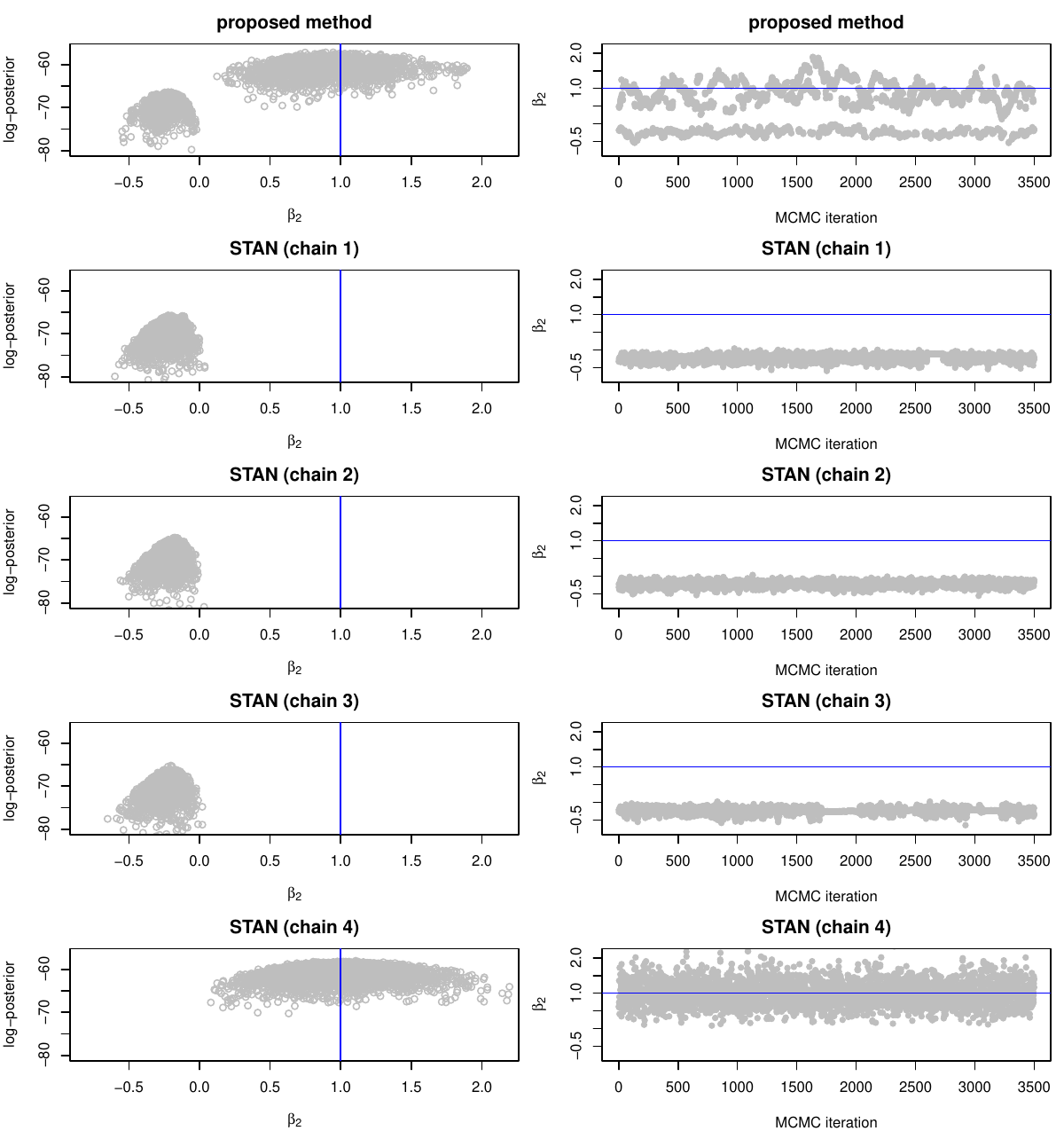}
\caption{\label{fig:stan_benchmark} Simulated values of $\beta_2$ versus the logarithm of the posterior density (up to a normalizing constant, left) and corresponding MCMC traces (right) with the proposed method (based on 4 heated chains) and 4 chains generated by STAN. The blue line denotes the true value of $\beta_2$. The first 1500 MCMC iterations have been discarded. }
\end{figure}

In this section we perform a comparison of the proposed method with STAN \citep{carpenter2017stan}, based on synthetic data generated by the proposed model (see the companion file \code{simulate_data.R}). For this purpose we have also coded our model in STAN programming language (see the companion file \code{cure_rate_model.stan}), for the special case where the promotion time is described by the Weibull distribution. The STAN implementation targets the marginal posterior distribution $\pi(\boldsymbol\theta|\boldsymbol y, \boldsymbol x)$ using the same prior distributions as the ones specified in Section \ref{sec:priors}. At first, we generate a synthetic dataset of $200$ observations based on the Weibull model. 
\begin{Schunk}
\begin{Sinput}
R> library("rstan")
R> library("pracma")
R> source("simulate_data.R")
R> truePars <- c(-0.05, 1, 0.8, 1, 2, -1, 1) 
R> n <- 200
R> myData <- sim_fotis_model(n = n, truePars = truePars, ab = 0.45, seed = 123)           
R> myData <- as.data.frame(myData)
\end{Sinput}
\end{Schunk}
The true values of the parameter vector is given in \code{truePars} in the following order: $\gamma$, $\lambda$, $\alpha_1$, $\alpha_2$, $\beta_0$, $\beta_1$, $\beta_2$. We generate an MCMC sample based on the proposed method for 5000 iterations and 4 heated chains, according to a Weibull model. 
\begin{Schunk}
\begin{Sinput}
R> mcmc_cycles <- 5000; nChains <- 4; nCores <- 1
R> set.seed(555, kind = "L'Ecuyer-CMRG")
R> start.time <- Sys.time()
R> fit_weibull <- cure_rate_MC3(Surv(Y, Censoring_status) ~ Covariate1 + Covariate2,
+  	data = myData, nChains = nChains, mcmc_cycles = mcmc_cycles,
+  	nCores = nCores, promotion_time = list(family = 'weibull'),
+  	verbose = FALSE)
\end{Sinput}
\begin{Soutput}
20 MCMC cycles required 0.3 secs. Expect a total run-time of: 76.21 secs. 
\end{Soutput}
\begin{Sinput}
R> end.time <- Sys.time()
R> time.taken <- end.time - start.time
\end{Sinput}
\end{Schunk}
Next, we run STAN using four chains. Note that we use the same set of (random) starting values as the one used in our method (these values are returned as output in the \code{fit_weibull} object in the \code{initial_values} entry. 
\begin{Schunk}
\begin{Sinput}
R> data_list <- list(
+    N = length(myData$Y),
+    y = myData$Y,
+    x1 = myData$Covariate1,
+    x2 = myData$Covariate2,
+    delta = myData$Censoring_status
+  )
R> inits <- vector("list", length = nChains)
R> for(i in 1:nChains){
+  	inits[[i]] <- vector("list", length = 7)
+  	names(inits[[i]]) <- c("gamma", "lambda", "alpha1", "alpha2", 
+  		"beta0", "beta1", "beta2")
+  	for(j in 1:7){inits[[i]][[j]] <- fit_weibull$initial_values[j,i]}
+  }
R> start.time <- Sys.time()
R> fit <- stan(file = "cure_rate_model.stan", data = data_list, init = inits,
+             iter = mcmc_cycles, chains = nChains, seed = 1)
R> end.time <- Sys.time()           
R> time.taken2 <- end.time - start.time
\end{Sinput}
\end{Schunk}
Figure \ref{fig:stan_benchmark} displays the sampled values of $\beta_2$\footnote{We have chosen the specific parameter to illustrate the results due to the fact that the multimodality of the posterior MCMC draws is vividly displayed.} versus the corresponding values of the posterior density. We conclude that three out of four chains in STAN remain trapped within a minor mode. It is evident that the four chains generated by STAN failed to mix. On the other hand, the proposed method moves freely around the posterior surface, constantly switching between the main and the minor mode.  The time for the proposed method is 83 seconds, while the time for STAN is 429 seconds.

\subsection{Comparison with the mixture cure model}
\label{sec:mixture}

We compare the proposed model against the \pkg{mixcure} \proglang{R} package \citep{mixcure,peng2021cure} which performs frequentist inference in the mixture cure rate model via the Expectation-Maximization algorithm. For this purpose we generated synthetic data from the mixture cure rate model using two covariates. We used the default logit link function in the mixture cure model, which is different than the exponential link function used in our model. Obviously, the models are different so there is no point in comparing the point estimates between the two approaches. However, since both approaches report an estimate of the cured probability for each censored item, we are focusing on the ability of each method to classify subjects as cured or not. 

\begin{figure}[ht]
\centering
\includegraphics{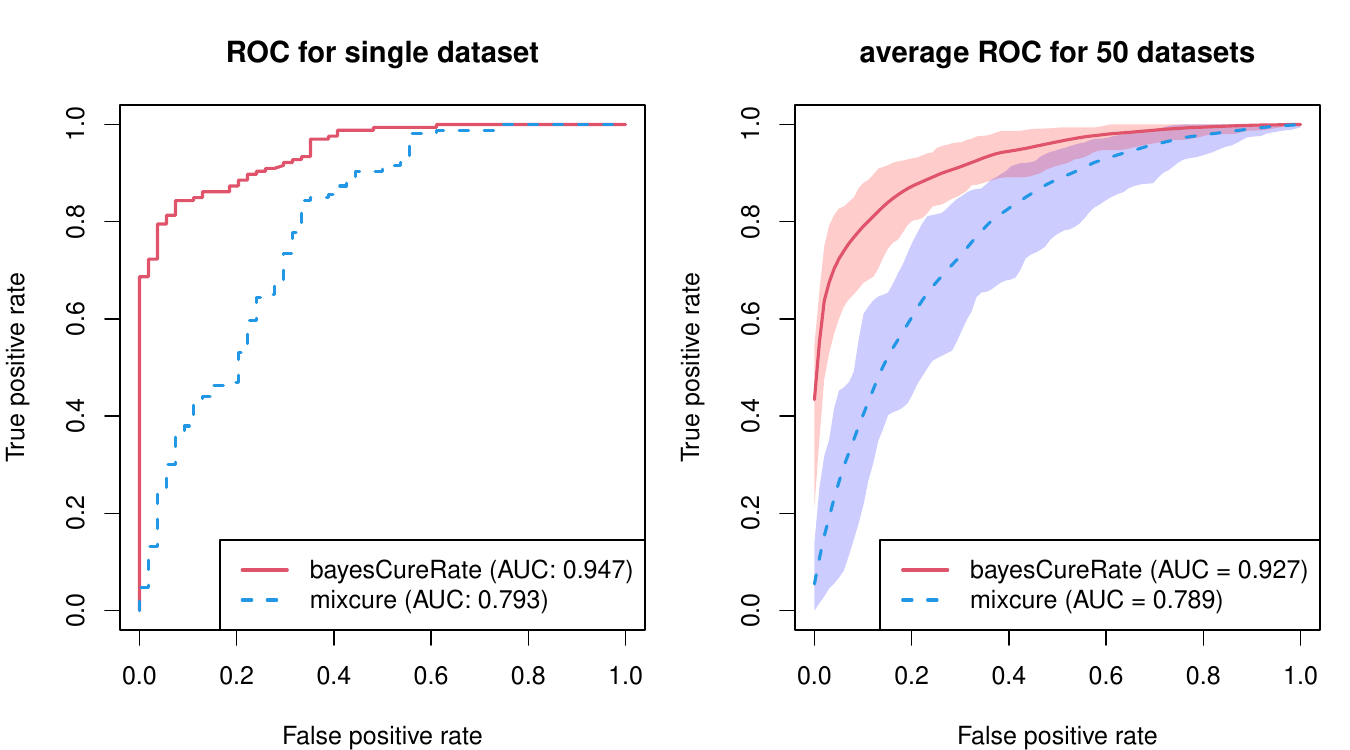}
\caption{\label{fig:mixcure_benchmark} Left: ROC curves for a single simulated dataset generated by the mixture cure rate model. Right: Average ROC curves with $95\%$ confidence bands for 50 synthetic datasets generated by the mixture cure rate model. }
\end{figure}

At first we generate a synthetic dataset from the mixture cure rate model (see the companion file \code{"sim_mixcure.R"}) consisting of 600 observations with two covariates. 
\begin{Schunk}
\begin{Sinput}
R> library("mixcure")
R> source("sim_mixcure.R")
R> #the set of true parameters
R> truePars <- c(1, 1, 1, 1, -1)
R> #sample size
R> nn1 <- 600
R> #simulation
R> newdata1 <- sim_model_logit (nn1, truePars, ab = 0.15, ranunL = -2, 
+    ranunU = 2, seed = 1)
R> newdata1 <-  as.data.frame(newdata1)
R> # true latent cure-status among the censored objects
R> # 1 = cured, 0 = susceptible
R> status1 <- newdata1$cured_status[which(newdata1$Censoring_status == 0)]
R> table(status1)
\end{Sinput}
\begin{Soutput}
status1
  0   1 
 54 166 
\end{Soutput}
\end{Schunk}
There are 220 censoring times among the 600 observations. The proportion of cured items within the censored items is equal to 75.5$\%$, that is, 166 cured subjects in total. At first we fit the mixture cure rate model according to the EM implementation in the \pkg{mixcure} package, using the Weibull distribution. 
\begin{Schunk}
\begin{Sinput}
R> model_mix_cure <- mixcure(Surv(Y, Censoring_status) ~ 1, ~ Covariate1+Covariate2, 
+            lmodel = list(fun = "survreg", dist = "weibull"), 
+            data = newdata1, savedata = TRUE)
R> cureprob1 <- predict(model_mix_cure, newdata1, 1)$cure[,2]
R> mixcure_model1 <- cureprob1[which(newdata1$Censoring_status == 0)]
\end{Sinput}
\end{Schunk}
The estimated cure probability per cencored subject is stored in \code{mixcure_model1}. 
Next, we run the proposed method using the Weibull distribution as well. 
\begin{Schunk}
\begin{Sinput}
R> mcmc_cycles <- 10000; nChains <- 4; nCores <- 1
R> set.seed(10, kind = "L'Ecuyer-CMRG")
R> run_WEI <- cure_rate_MC3(Surv(Y, Censoring_status) ~ Covariate1+Covariate2,
+    data = newdata1, nChains = nChains, mcmc_cycles = mcmc_cycles,
+    nCores = nCores, promotion_time = list(family = 'weibull'), verbose = FALSE)
\end{Sinput}
\begin{Soutput}
20 MCMC cycles required 0.51 secs. Expect a total run-time of: 253.74 secs. 
\end{Soutput}
\begin{Sinput}
R> run_wei_sum1 <- summary(run_WEI, burn = 3000) 
R> bayescure_model1 <- run_wei_sum1$latent_cured_status
\end{Sinput}
\end{Schunk}
The estimated cure probability per cencored subject is stored in \code{bayescure_model1}. Figure \ref{fig:mixcure_benchmark} (left) displays the ROC curve for both methods. We conclude that the proposed method (bayesCureRate) outperforms the implementation in the \pkg{mixcure} package, despite the fact that we are simulating data from the model in the latter package. Next, we have replicated the previous simulation 50 times, using the same parameter setup. The averaged ROC curve is displayed in Figure \ref{fig:mixcure_benchmark} (right), along with $95\%$ confidence intervals. We conclude once again the superior classification performance of the proposed method.

\end{document}